\documentclass[twocolumn]{aastex631}

\shorttitle{GULP Results: Stellar Hierarchy}
\shortauthors{Meena et al.}

\graphicspath{{./}{figures/}}

\usepackage{color}
\usepackage{comment}
\usepackage[normalem]{ulem}
\usepackage{natbib}
\usepackage{amsmath}
\defcitealias{Sabbi2025}{Paper~I}
\newcommand{\paperI}{Sabbi et al. (in prepration)} 

\definecolor{malachite}{rgb}{0.01, 0.8, 0.24}

\usepackage{soul,xcolor}
\setstcolor{red}

\begin{document}
\title{GULP II: Hierarchical Distribution and Evolution of Young Stellar Structures in NGC~4449}

\correspondingauthor{Beena Meena}
\email{bmeena@stsci.edu}
\author[0000-0001-8658-2723]{Beena Meena}
\affiliation{Space Telescope Science Institute, 3700 San Martin Drive, Baltimore, MD 21218, USA}

\author[0000-0003-2954-7643]{Elena Sabbi}
\affiliation{Space Telescope Science Institute, 3700 San Martin Drive, Baltimore, MD 21218, USA}

\author[0000-0002-6091-7924]{Peter Zeidler}
\affiliation{AURA for the European Space Agency, ESA Office, STScI, 3700 San Martin Drive, Baltimore, MD 21218, USA}

\author[0000-0002-1723-6330]{Bruce G. Elmegreen}
\affiliation{Katonah, New York USA 10536}

\author[0000-0002-1722-6343]{Jan J. Eldridge}
\affiliation{Department of Physics, University of Auckland, Private Bag 92019, Auckland, New Zealand}

\author{Varun Bajaj}
\affiliation{Space Telescope Science Institute, 3700 San Martin Drive, Baltimore, MD 21218, USA}

\author[0000-0002-5581-2896]{Mario Gennaro}
\affiliation{Space Telescope Science Institute, 3700 San Martin Drive, Baltimore, MD 21218, USA}

\author[0000-0001-5171-5629]{Anna Pasquali}
\affiliation{Astronomisches Rechen-Institut, Zentrum f\"{u}r Astronomie der Universit\"{a}t Heidelberg, M\"{o}nchhofstra{\ss}e 12-14, D-69120 Heidelberg, Germany}

\author[0000-0002-1392-3520]{Debra M. Elmegreen}
\affiliation{Vassar College, Dept. of Physics \& Astronomy, Poughkeepsie, NY 12604}

\author[0000-0002-0560-3172]{Ralf S.\ Klessen}
\affiliation{Universit\"{a}t Heidelberg, Zentrum f\"{u}r Astronomie, Institut f\"{u}r Theoretische Astrophysik, Albert-Ueberle-Str.\ 2, 69120 Heidelberg, Germany}
\affiliation{Universit\"{a}t Heidelberg, Interdisziplin\"{a}res Zentrum f\"{u}r Wissenschaftliches Rechnen, Im Neuenheimer Feld 225, 69120 Heidelberg, Germany}
\affiliation{Harvard-Smithsonian Center for Astrophysics, 60 Garden Street, Cambridge, MA 02138, U.S.A.}
\affiliation{Elizabeth S. and Richard M. Cashin Fellow at the Radcliffe Institute for Advanced Studies at Harvard University, 10 Garden Street, Cambridge, MA 02138, U.S.A.}

\author[0000-0002-0806-168X]{Linda J. Smith}
\affiliation{Space Telescope Science Institute, 3700 San Martin Drive, Baltimore, MD 21218, USA}

\author[0000-0001-7746-5461]{Luciana Bianchi}
\affiliation{Dept. of Physics \& Astronomy, The Johns Hopkins University, 3400 N. Charles St., Baltimore, MD
21218, US}

\author[0000-0001-8289-3428]{Aida Wofford}
\affiliation{Instituto de Astronom\'{i}a, Universidad Nacional Aut\'{o}noma de M\'{e}xico, Unidad Acad\'{e}mica en Ensenada, Km 103 Carr. Tijuana-Ensenada, Ensenada 22860, M\'{e}xico}

\author[0009-0003-6044-3989]{Pietro Facchini}
\affiliation{Astronomisches Rechen-Institut, Zentrum f\"{u}r Astronomie der Universit\"{a}t Heidelberg, M\"{o}nchhofstra{\ss}e 12-14, D-69120 Heidelberg, Germany}

\author[0000-0001-8608-0408]{John S. Gallagher III}
\affiliation{Department of Astronomy, University of Wisconsin-Madison, 475 North Charter St. Madison, WI 53706 USA}

\author[0000-0002-5189-8004]{Daniela Calzetti}
\affiliation{Department of Astronomy, University of Massachusetts Amherst, 710 North Pleasant Street, Amherst, MA 01003, USA}

\author[0000-0002-1891-3794]{Eva K. Grebel}
\affiliation{Astronomisches Rechen-Institut, Zentrum f\"{u}r Astronomie der Universit\"{a}t Heidelberg, M\"{o}nchhofstra{\ss}e 12-14, D-69120 Heidelberg, Germany}

\author[0000-0002-8192-8091]{Angela Adamo}
\affiliation{Department of Astronomy, The Oskar Klein Centre, Stockholm University, AlbaNova, SE-10691 Stockholm, Sweden}

\collaboration{24}{(GULP Collaboration)}

\begin{abstract}
We investigate the hierarchical distribution and evolution of young stellar structures in the dwarf starburst galaxy NGC~4449 using data from the GULP survey. By analyzing the spatial distribution of field stars younger than 100 Myr, we identify large-scale stellar complexes and substructures using HDBSCAN --a density-based clustering algorithm-- and trace their evolution over time. While comparing these stellar structures in different regions of the galaxy, we find that the central bar-like region shows a clear expansion of the structures within the first $\sim$ 60 Myrs, while the arm-like structure in the NE shows no discernible trend, possibly due to external perturbations from tidal interactions with a neighboring galaxy. An age-dependent two point correlation function (TPCF) analysis shows that young stars exhibit a strong hierarchical distribution, with clustering strength decreasing over time. The power-law slope of the TPCF, which starts at $\alpha \sim 0.65$ for stars younger than 5 Myr, shows a slight decline to $\alpha \sim 0.4$ for stars older than 50 Myr, though it does not reach a completely flat (random) distribution. This trend indicates a subtle weakening of structural hierarchy among young ($<$100 Myr) stars, which is primarily driven by internal stellar motions. Future work will extend this analysis to the remaining 26 galaxies in the GULP survey to better constrain the role of the galactic environment in shaping the hierarchical evolution of young stellar populations.
\end{abstract}


\section{Introduction}\label{sec:intro}

Star formation spans multiple scales, from giant molecular clouds to individual stars \citep{Lada2003}. Observations show that star-forming regions follow a nested structure, where larger structures break into smaller, denser groupings \citep{Bastian2007, Gouliermis2015}. This multi-scale organization produces various types of stellar systems such as associations, complexes, and aggregates, which create stellar-hierarchy \citep{Elmegreen2008, Bianchi2014, Gouliermis2015, Gouliermis2018} across the galaxy. These hierarchical patterns arise from dynamic processes within the interstellar medium (ISM), including gravity, supersonic turbulence, and feedback, as well as external influences such as galactic rotation, shear, and velocity perturbations introduced by structures like spiral arms, bars, and tidal interactions \citep{Elmegreen2004, Scalo2004, MacLow2004, McKee2007, Klessen2016, Vazquez-Semadeni2017, Girichidis2020}.

Stars typically form in clustered environments, but they do not remain in their birth place indefinitely. Turbulence within the ISM plays a critical role in shaping the hierarchical structure of star-forming regions by fragmenting molecular clouds. Once stars have formed, processes such as the cluster velocity dispersion, cloud destruction from stellar feedback through radiation and winds, supernovae, tidal forces, and galactic shear contribute to their eventual dispersal into the galactic field. Over time, clusters and stellar associations either undergo rapid disruption independently of their mass \citep{Fall2005,Chandar2006,Parker2014} or gradually dissolve into the field of a galaxy with lower mass ones dissolving faster \citep{Schmeja2006,Pellerin2007,bastian2012, Krumholz2019}. Evidence shows that environmental dynamical conditions might also play a role in the disruption rates \citep[e.g.,][]{messa2018, linden2022}, though the precise mechanisms and timescales governing this dispersal remain a topic of active investigation. 

Observational studies have employed various methods such as Kernal Density Estimation \citep[e.g.,][]{Gouliermis2017, Ksoll2021, Larson2023}, Friends of Friends  \citep[e.g.,][]{Pellerin2012, Drazinos2013, VargasSalazar2020, Chi2023}, Path Linkage Criteria \citep[e.g.,][]{,Bianchi2014,Rodriguez2016, Rodriguez2020}, and other clustering algorithms to identify multi-scale stellar structures in nearby star-forming galaxies (such as minimum spanning trees and dendrograms, e.g.\ \citealt{Cartwright2004}, \citealt{Gouliermis2010}, or \citealt{Rodriguez2019}). Most popularly, the two point correlation function (TPCF) has been employed often to  investigate the spatial distribution of stars and star-clusters in nearby galaxies. 

Using the TPCF on star-clusters in nearby galaxies, several studies \citep{Zhang2001,Scheepmaker2009,Gouliermis2014, Grasha2015,Grasha2017a,Menon2021} have consistently shown that young, massive star-clusters tend to be arranged in a clustered configuration, often associated with molecular cloud complexes. However, this hierarchical arrangement diminishes as the star-clusters age. Using the connection between the age difference and spatial separation among young star-clusters in local galaxies, \citet{Efremov1998} and \citet{Grasha2017b} reported that star formation is hierarchical in both space and time, where star-clusters of similar age form close together. They suggest that turbulence plays a key role in driving this hierarchy and that the maximum size of star-forming regions may also be influenced by the galaxy's shear. Recently, using Far-UV observation of nearby galaxies and TPCF, \citet{Shashank2025} show that hierarchies in young star-forming clumps persists only up to a certain physical scale and does not extend across the full galaxy. Moreover, this hierarchy disperses over a timescale of 10–15 Myr.
Earlier, theoretical studies by \citet{Elmegreen2018} examined the dispersal of young stellar hierarchies, proposing that the observed time decay of the power-law in the TPCF may result not just from random stellar motions or galactic shear, but also from the superposition of different hierarchies formed by multiple generations of stars.

Tracing individual stars allows us to understand how stellar structures evolve and disperse over time. Observations of young stellar populations in nearby galaxies suggest that their spatial distribution reflects the fractal structure of their parent molecular clouds, shaped by turbulence and self-gravity \citep{Rodriguez2020}. Earlier studies of the Magellanic Clouds have shown that stars are formed in highly sub-structured environments, with the spatial distribution gradually approaching that of the field population on timescales comparable to a galaxy’s crossing time \citep{Gieles2008, Bastian2009}. Similarly, using time-dependent clustering of the young stellar population in the ring galaxy NGC 6503, \cite{Gouliermis2015} observed a transition from clustered to dispersed young populations over $\sim$60 Myr, driven by turbulence and shear. The study highlighted hierarchical star formation within the ring, with shear as the primary driver of turbulence based on rotational velocity differences at the edges of the ring. Building on these studies of young stellar populations, in this work, we examine spatial and temporal distribution of bright young massive stars in the dwarf galaxy NGC~4449. Our goal is to understand how stars migrate from their birthplaces and populate the galaxy's field, and to determine whether these dispersal processes are universal across different galactic environments.

\begin{figure}[htbp!]
\includegraphics[width=0.46\textwidth]{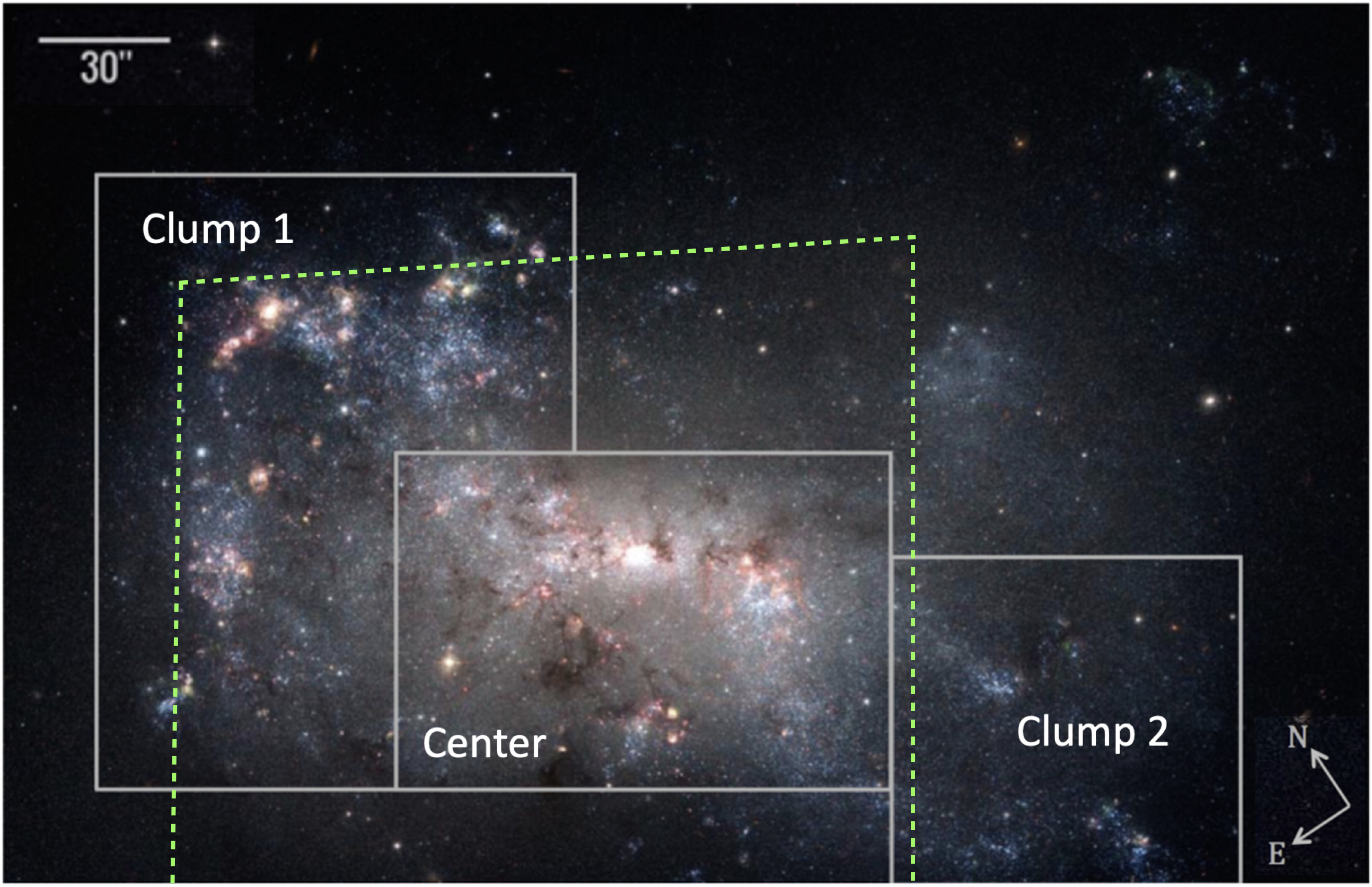}
\caption{A color-composite optical image of NGC 4449 using multiple HST filters (blue: F435, green: F555W, red: F814W and H$\alpha$ in magenta). This image is adopted from \cite{Sacchi2018} (Figure~12), who derived the SFH in the three sub-regions `Clump1', `Center' and `Clump2'. The boundaries of these regions are shown in white lines. In this work, we use the stars with valid photometry in F275W image (footprint shown in dashed light-green lines), which covers only Clump1 and Center, thus excluding the older `Clump2' region from the analysis.} 
\label{fig:NGC4449_regions}
\end{figure}

NGC~4449 is a nearby irregular Magellanic-type dwarf galaxy located at a distance of 4.01 Mpc \citep{Sabbi2018}. The line-of-sight (LOS) inclination and the position angle (PA) of the galactic disk are $=$64\arcdeg~and $=$ 45\arcdeg, respectively \citep{Hunter2005}. This galaxy is classified as a starburst \citep{McQuinn2010, Whitaker2012, Lelli2014, Calzetti2018}, known for its active star-forming regions scattered throughout the galaxy. The star formation rate (SFR) is approximately 0.5~M$_{\odot}$~yr$^{-1}$ \citep{Lee2009}, with the most recent episode of star formation likely occurring between 5 and 10 million years ago \citep{McQuinn2012,Sacchi2018, Cignoni2019}.

DDO 125, the closest neighbor to NGC~4449, lies at a projected distance of 40 kpc \citep{Theis2000}. Several studies \citep{Hunter1998, Theis2000,Martinez-Delgado2012, Rich2012, Ai2023} have provided evidence of a minor merger and interactions with its neighboring galaxies, which likely triggered the intense starburst in this galaxy.

Using synthetic color-magnitude diagrams (CMDs) and stellar evolution models, \cite{Sacchi2018} traced the star-formation history (SFH) in different regions of the inner part of the galaxy (consisting the bright and denser population) by splitting it into three sub-regions designated as `Center' , `Clump 1' and `Clump2' as shown in Fig.~\ref{fig:NGC4449_regions}. The Center is densely populated with bright young stars and contains a well-defined bar \citep{Hill1998}. This region had an intense and prolonged episode of SF, with two major peaks around 10 and 100 Myr ago. Clump1 is located near the northern part of the galaxy, and contains the two stellar streams and an elongated structure, likely formed by an interaction or merger \citep{Cignoni2019}. Clump2 is situated southwest of the center and, in contrast to Clump1, exhibits a lower star formation rate. We will later discuss the implications of these regions in relation to their hierarchical properties. Since our observational field of view (FoV) only covers the Center and Clump1, we will focus on how the galaxy's environment has shaped the spatial distribution of stars in these two regions.

This paper is organized as follows. In \S\ref{sec:obs}, we provide an overview of the GULP survey observations and describe the data used in this study. In \S\ref{sec:analysis}, we identify stellar structures in NGC~4449 using density-based clustering (\S\ref{subsec:steller_structures}) and quantify stellar hierarchy through two-point correlation function (TPCF) analysis (\S\ref{subsec:tpcf} and \ref{subsec:mcmc}), followed by the associated observational results. In \S\ref{sec:discussion} we discuss the physical implications of these findings and explore the processes driving the evolution and dispersal of stellar hierarchy in this galaxy. Finally, in \S\ref{sec:summary}, we summarize our conclusions.

\section{Observations and Data}\label{sec:obs}

\begin{figure*}[htbp!]
\includegraphics[width=0.98
\textwidth]{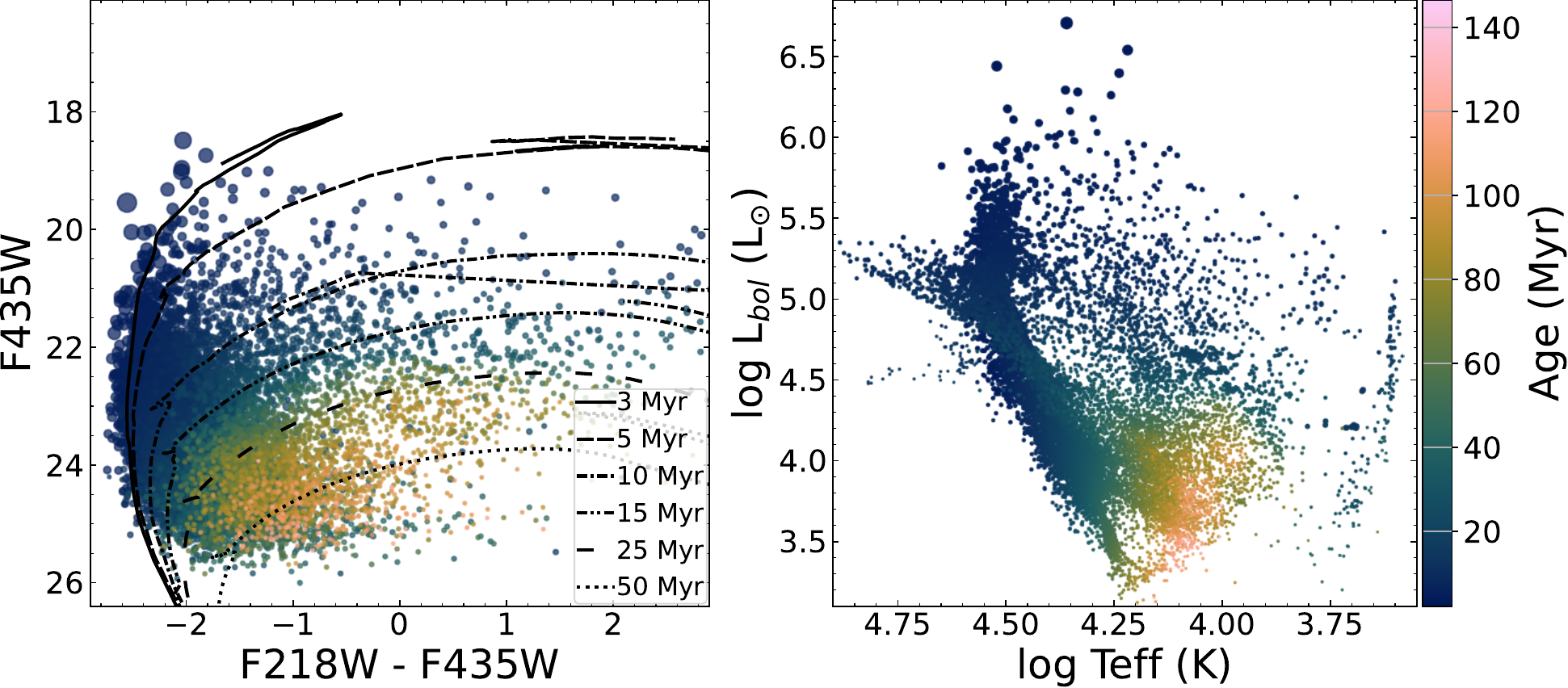}
\caption{Left: A CMD using observed F435W magnitude (y axis) vs F218W $-$ F435W color (x axis). Padova isochrones for metallically ($Z$) = 0.004, distance modulus = 28.02, $E(B-V)$ = 0.07 mag \citep{Sabbi2018} and extinction law in \cite{Gordon2016} with $R(V)=$~2.7 and $f_{A}=$~0 are overplotted in black curves of different linestyles for ages 3 Myr (solid), 5 Myr (densly dashed), 10 Myr (dashdotted), 15 Myr (dashdotdotted), 25 Myr (loosely dashed) and 50 Myr (dotted). Right: HR diagram using the mean-fit parameters obtained from BPASS model. The x axis shows the effective temperature (in log) and y axis is the bolometric luminosities (in log) for the the stars (or primary star for the binary systems). In both figures, the detected sources are color coded by age. The color bar is shown on the right. Additionally, the size of the points is proportional to the mass of the primary star (ranging from $\sim$1 $M_\odot$ to $\sim$175 $M_\odot$). With our GULP observations (aided by F150LP), we have a strong constraint on stellar temperatures up to 50,000 K. The points to the left likely indicate Wolf-Rayet stars. However, the parameters shown here reflect the mean fit, not the best fit—hence the discrepancy. More details will be discussed in Paper I.}
\label{fig:HR_cmd}
\end{figure*}

\begin{figure*}[htbp!]
\includegraphics[width=\textwidth]{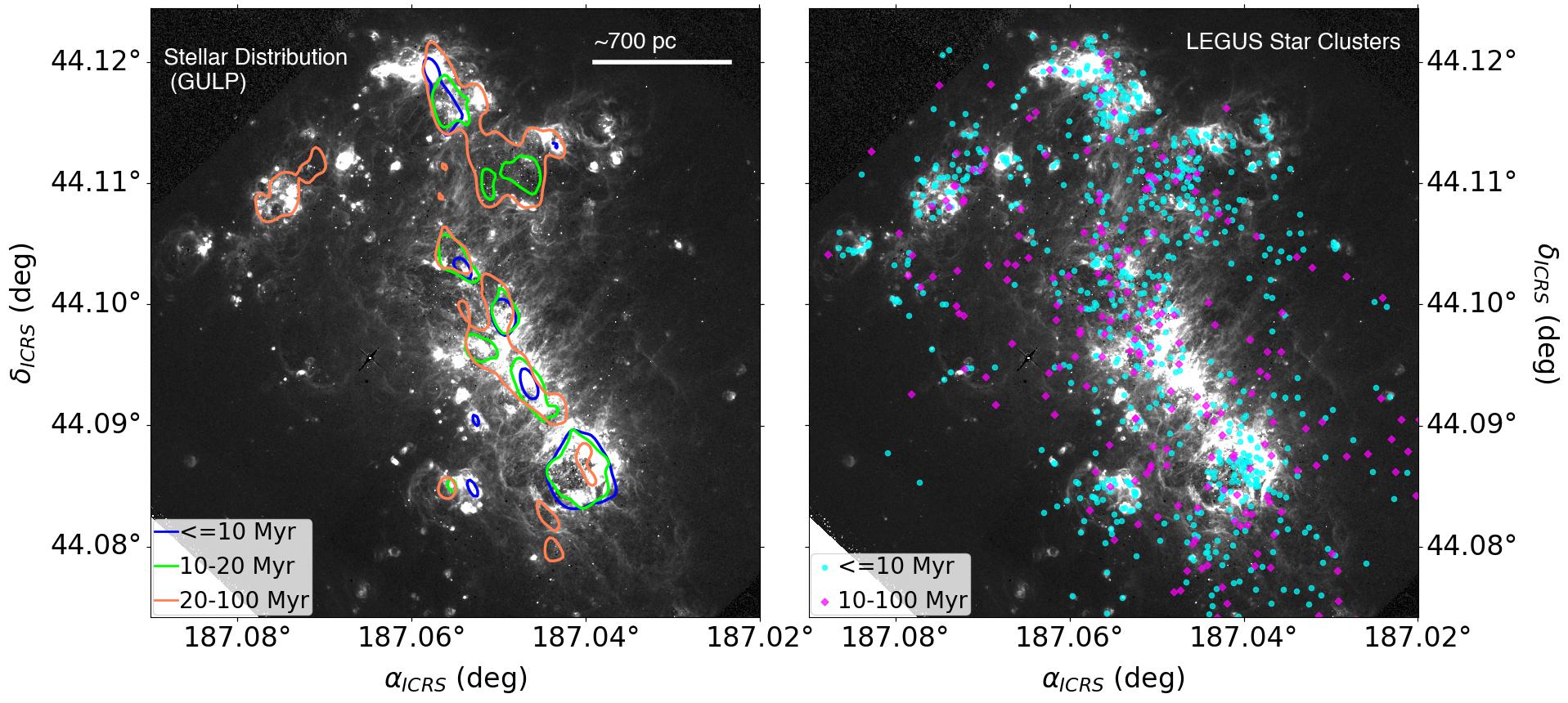}
\caption{Left: Continuum (F814W) subtracted narrowband (F668N) H$\alpha$ image of NGC~4449, overlaid with  (KDE) contours representing different stellar age groups. The contours illustrate the spatial distribution of stars younger than 10 Myr (blue), between 10 and 20 Myr (green), and between 20 and 100 Myr (orange). The KDE contours are plotted for regions with densities exceeding 3$\sigma$ above the background, using a kernel density bandwidth of approximately 60 pc. The right panel shows the spatial distribution of star-clusters identified in LEGUS catalogs \citep{Cook2023}. It is important to note that the larger FoV of the LEGUS observations results in a slightly broader distribution of star-clusters compared to the field stars observed in the GULP survey. In both images North is up, East is on the left and 1\arcsec~$\approx$ 1.9 pc.}
\label{fig:Ha_image}
\end{figure*}

The observations used in this work are part of the Galaxy UV Legacy Project (GULP) Survey. GULP (PI: Elena Sabbi, GO-16316) is a large treasury program designed to characterize the properties of resolved massive stars, OB associations, and the field star populations of 27 nearby star-forming galaxies using high resolution capabilities of Hubble Space Telescope (HST) in near and far ultraviolet (NUV and FUV). The goal of this project is to constrain the high-mass end of the initial mass function (IMF) and the extinction curves (particularly the characterization of the UV bump) and to derive robust age estimates for the young, massive stars. 

In the FUV, we utilize the Solar Blind Channel (SBC) of the Advanced Camera for Surveys (ACS) in the F150LP filter (pixel scale $\sim$ 0.03$\arcsec$) and in the NUV we use the UVIS channel of the Wide Field Camera 3 (WFC3), in the F218W filter (pixel scale $\sim$ 0.04$\arcsec$). We combine these observations with the archival longer wavelength HST observations available from Legacy ExtraGalactic UV Survey (LEGUS, \citealp{Calzetti2015} in the filters F275W, F336W, F438W, F555W, and F814W) surveys, which give us a multi-waveband dataset spanning from the Far UV to the I band.

A detailed overview of the GULP survey and descriptions of of the observations, data reduction, drizzling and photometry is provided in \paperI (hereafter Paper-I).

In order to derive the physical properties of the stars (including masses, ages, temperatures, luminosities), we employ stellar population synthesis models using the publicly available Binary Population and Spectral Synthesis v2.2 code suite (BPASS, \citealp{Eldridge2017}, \citealp{Stanway2018}) for binary stellar population (see Paper-I). We perform the SED fitting only on the sources that were detected in at least five broadband filters (while ensuring detection in the F275W filter), with photometric errors below 0.1 mag in each band. Additionally, we always use F555W and F814W data, which have a larger FoV than F150LP and are comparable to F275W, to account for dust extinction and stellar masses.

For the modeling, we adopt the extinction law described by \cite{Gordon2016} with $E(B-V)$~$=$~0.07 mag (\citealp{Sabbi2018}, after subtraction of the foreground Milky Way extinction), $R(V)=$~2.7 and $f_{A}=$~0 (Paper-I). Here, $R(V)$ and $f_{A}$ correspond to the dust average grain size and the dust mixer coefficients respectively (see \citealp{Gordon2016} for their definitions). These extinction parameters were obtained by superimposing Padova isochrones \citep{Bressan2012, Pastorelli2019, Pastorelli2020} over various combinations of (FUV to I band) color-magnitude diagram (CMD)s. An example of such a CMD using the F435W vs F218W$-$F435W with overplotted isochrones for different ages is shown in Figure~\ref{fig:HR_cmd} (left).

We compare each source SED against every timestep from every stellar model in the BPASS model set at a metallicity of $Z=0.004$ \citep{Sabbi2018}. If the model SED is more luminous than the observed SED we allow for the inclusion of extra dust as one of our fitting parameters. To determine which model is the best fit to the observed SED, we calculate the quality of each fit by using the model and observed magnitudes and how well they agree within the photometric errors. This probability of the fit is multiplied by the model weight, calculated from  the IMF \citep{2001MNRAS.322..231K}, the initial binary population distribution \citep{2017ApJS..230...15M} and the time the star spends with the model SED. This allows us then to find the best-fitting model, the mode fit, as the model with the highest probability, and allows us to calculate a mean fit that identifies the most like parameter space for each source. This method has been adapted from the one outlined in \citet{2011MNRAS.411..235E}. The result is that for each source we have an initial mass, age and $A_{\rm V}$\footnote{We allow for varying extinction to each source given that Wolf-Rayet stars and red supergiants can produce large amounts of dust. While some sources may be embedded in local dust not considered within the whole galaxy extinction that has been derived.} and if it is a binary star an initial mass ratio, initial orbital period and nature of the companion (i.e., another main sequence star or compact remnant). These then allow us to derive the current masses, stellar age, luminosity, temperature, gravity and other parameters as required. Figure~\ref{fig:HR_cmd} (right) shows a Hertzsprung-Russell (HR) diagram using the mean fit model outputs from BPASS. The corresponding color-magnitude diagram (CMD; left panel), where stars are color-coded by their ages derived from BPASS SED fitting, also demonstrates good agreement with the ages inferred from the Padova isochrones. More detailed discussions of NGC 4449 extinction curves and nuances of different SED fitting outcomes (including single and binary models) and various stellar populations is provided in Paper-I.

In this work, we focus source detection in the UV (using F275W) to enhance sensitivity to young, massive stars and to break the temperature–age degeneracy—particularly using the F150LP and F218W filters. This also improves constraints on extinction. Having required a UV detection, our look-back time is reduced to a few hundred Myr, and our sample start to suffer for incompleteness already for populations older than $\sim$60–100 Myr (see \S\ref{sec:discussion}).

Figure~\ref{fig:Ha_image} shows a kernel density estimate (KDE) distribution of stars grouped for ages $<$10 Myr, 10-20 Myr and 20-100 Myr. As expected, the spatial distribution of the young massive stars from the GULP observations shows a strong correlation with the distribution of ionized gas, traced by the continuum-subtracted H$\alpha$ image. Notably, evidence of recent star formation is observed throughout the galaxy, including in the `Clump1' and `Center' regions \citep{Sacchi2018}. However, the KDE contours indicate a clear migration of star formation activity over time from North to South. In the North, the stellar distribution is dominated by stars older than 20 Myr, whereas the central region exhibits a significant concentration of younger ($<$ 20 Myr) stars, reflecting the a more recent star formation activity towards the center of the galaxy. This trend highlights the evolving nature of star formation within NGC 4449.

Additionally, the right panel in Figure~\ref{fig:Ha_image} presents the distribution of star-clusters identified in the LEGUS survey \citep{Cook2019, Cook2023} for comparison. This visualization highlights a strong concentration of star-clusters younger than 10 Myr, while older populations appear more dispersed. This distribution will be utilized later in \S\ref{subsec:tpcf} to compare the two-point correlation functions (TPCF) of field stars from the GULP observations with those of star-clusters from LEGUS. It is important to highlight that, in this paper, we use the terms ``clustering'' or ``clustered'' to refer to the grouping or spatial concentration of field stars, whereas ``star-cluster'' specifically denotes Class 1, 2, and 3 objects identified in the LEGUS catalog \citep{Cook2023}.

\section{Analysis and Observed Results}\label{sec:analysis}

\subsection{Deprojecting Stellar Separation}\label{subsec:deproject}

\begin{figure*}[ht!]
\includegraphics[width=\textwidth]{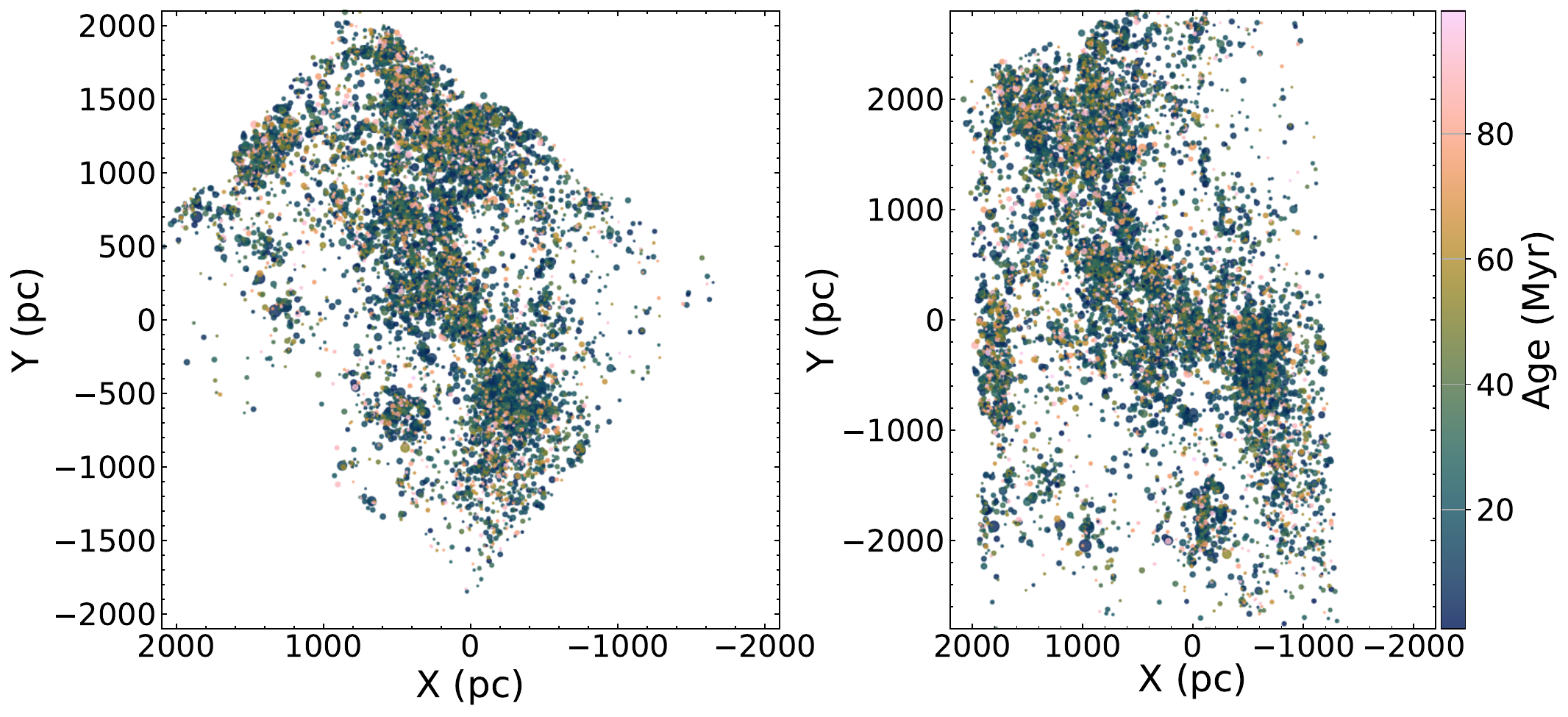}
\caption{The right panel shows the observed spatial distribution of the field stars and the left panel shows the de-projected positions of the stars obtained by rotating along the major axis and correcting for the inclination of the galaxy's disk. Here only the stars that are younger than 100 Myr are shown and the colorbar on the right shows their ages. X and Y $=$ $0$ (pc) indicates the adopted galaxy center. }
\label{fig:obs_deproject}
\end{figure*}

Deprojecting a galaxy in the sky is crucial to accurately determine the separation between stars and characterize the stellar structures and density distributions within a galaxy rest frame. 
In order to de-project the observed coordinates (x and y positions) of the stars, we first assume an axisymmetric flat rotating disk and rotate the positions of the stars in clockwise direction from the center of the galaxy by an angle $\phi$ using the following equations:
\begin{flalign}
    x' = x~\cos(\phi) + y~\sin(\phi), \\
    y' = -x~\sin(\phi) + y~\cos(\phi),
\end{flalign}\label{eq:deproject}
where $\phi$ is the PA of the major axis of the host galaxy and $x'$ and $y'$ are the PA corrected coordinates of the stars. We then correct these coordinates for the LOS inclination ($i$) using $x' = x'$ and $y' = y'/\cos(i)$, which gives us the de-projected distances of the stars from the center. For NGC~4449, we adopt $\phi$ = 45\arcdeg~and $i$~=~64\arcdeg~\citep{Hunter2005}.

Figure~\ref{fig:obs_deproject} shows the observed and the de-projected spatial distribution of the stars younger than 100 Myr. We will use these positions and ages of the field stars and star-clusters for further analysis. However it is to be noted this analysis remains limited by the unknown depth extent along the line of sight (in z direction), which could influence the de-projected spatial distribution of stars.

\subsection{Identifying Stellar Complexes and Substructures}\label{subsec:steller_structures}

To identify potential stellar structures across different age groups and uncover any structural correlations or evolutionary trends related to age, we utilize an unsupervised machine learning clustering algorithm. Specifically, we implement a density-based clustering algorithm - HDBSCAN (`Hierarchical Density-Based Spatial Clustering of Applications with Noise', \citealp{Campello2013, McInnes2017}) on the spatial distribution (X $\&$ Y positions from the center of the galaxy) of the field stars in different age bins ($<$5 Myr, 5-10 Myr, 10-20 Myr, 20-30 Myr, 30-50 Myr, and 50-100 Myr). By choosing narrower age ranges for younger stars and wider ranges for older stars, we aim to capture the structural behavior of the very young stellar population while ensuring sufficient statistical robustness for the older populations. This approach enables a detailed examination of the spatial distribution and clustering behavior across stellar populations of different ages.   

HDBSCAN offers advantages over traditional distance-based clustering algorithms like K-Means \citep{Jin2010} and Mean-shift \citep{comaniciu_mean_2002} by identifying clusters of varying sizes and shapes without any prior knowledge of the number of clusters, while managing noise and outliers, and revealing hierarchical structures within large spatial datasets. It is particularly effective in identifying structures in environment with variable density.

\begin{figure*}[htbp!]
\includegraphics[width=0.495\textwidth]{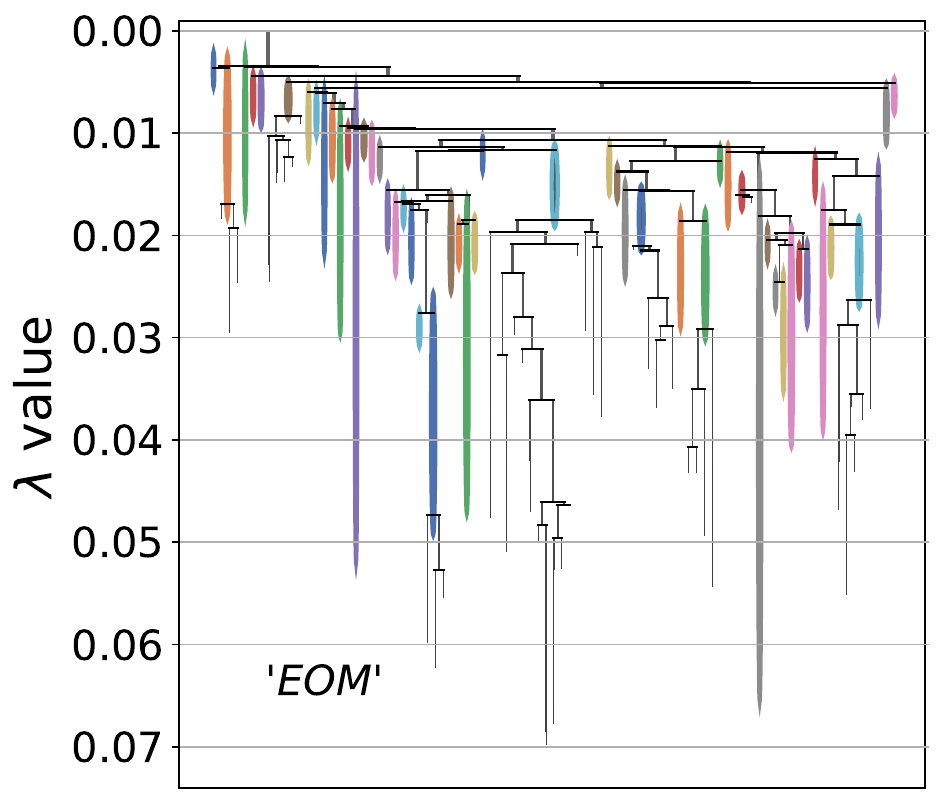}
\includegraphics[width=0.495\textwidth]{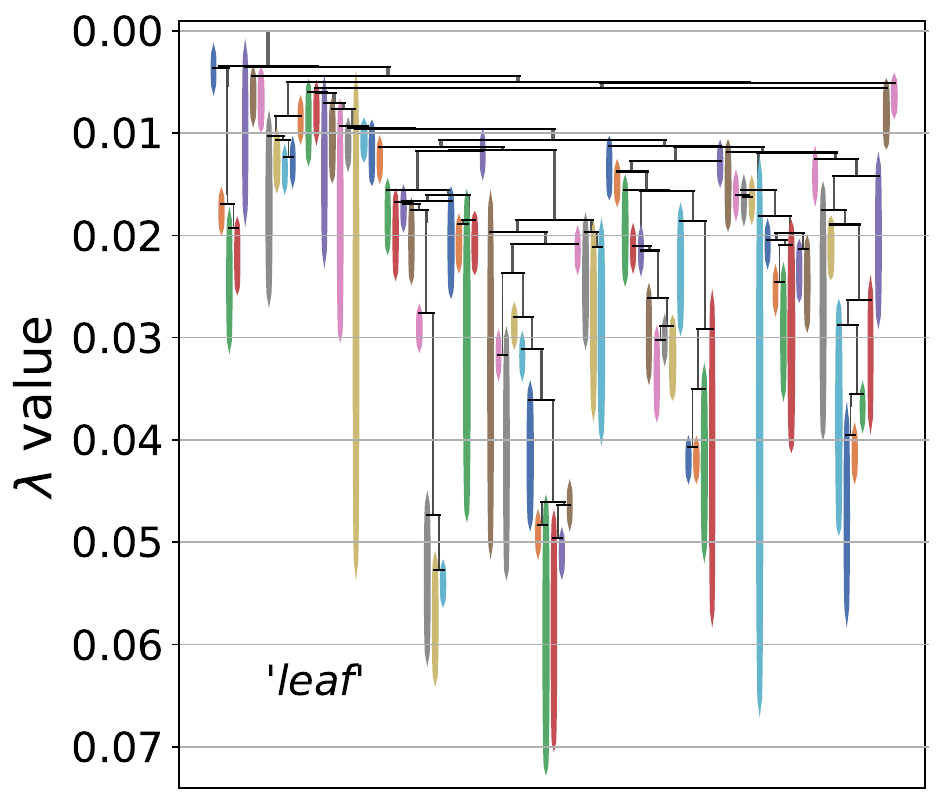}
\caption{The figure displays the condensed tree plot from HDBSCAN for the stellar population aged 10-15 Myr. In both panels, the black thin branches represent all potential cluster groupings, while the colored branches indicate the clusters selected by the `cluster selection method'. The left panel highlights the `EOM' clusters, and the right panel highlights the `leaf' clusters. The y-axis shows the $\lambda$ value, which increases with the cluster density. The `EOM' clusters are the most stable clusters, determined by the area under each branch. Stable clusters span a large vertical range on the tree, persisting across a broad range of $\lambda$ values, which indicates they are strong and reliable under different density environments. In contrast, the leaf clusters represent the most granular level of the hierarchy and are located at the tips of the branches, where the clusters can no longer be split into smaller subclusters.}
\label{fig:condensed_tree}
\end{figure*}

To automatically identify clusters with arbitrary shapes in a large dataset with spatially varying density, HDBSCAN constructs a hierarchical representation of clusters, often visualized through a Condensed Tree (see Figure~\ref{fig:condensed_tree}), which is similar to a dendrogram. The tree illustrates how clusters merge and split at different density levels, offering insights into the dataset's structure. HDBSCAN uses `cluster stability' to select the most reliable clusters, classified as `EOM' (excess of mass) clusters in the condensed tree. Cluster stability is derived from this tree by evaluating how long (in terms of $\lambda$, which is inverse of mutual reachability distance) a branch or cluster persists. A more stable cluster will survive for a wider range of lambda values before it either breaks apart or merges with another cluster. The area of each branch in the tree represents the stability of that cluster.

\begin{figure*}[htbp!]
\includegraphics[width=\textwidth]{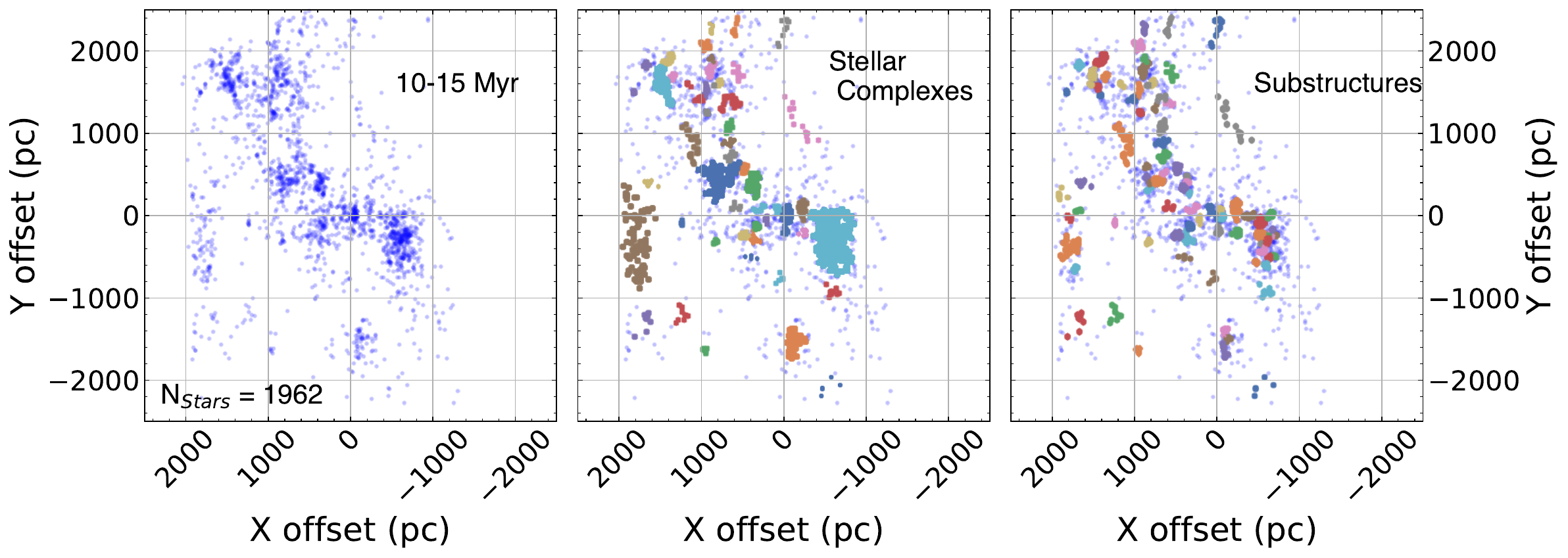}
\caption{This figure presents the spatial distribution of stars in the 10-15 Myr age group. The leftmost panel displays the overall spatial distribution of stars, represented by blue points. The middle panel highlights the `EOM' clusters in different colors, corresponding to the clusters identified in the condensed tree plot (left) where branches were highlighted in same colors. The rightmost panel displays the same stellar distribution with `leaf' clusters highlighted, representing finer substructures within the larger stellar complexes identified in the middle panel. The points that are in blue in the middle and rightmost panel are identified as `noise' by HDBSCAN and are not part of any stellar complexes/structures.}
\label{fig:age_cluster}
\end{figure*}

HDBSCAN offers two methods for selecting clusters from the condensed tree: the default `EOM' method, which identifies the most persistent and stable clusters as explained above, and the `leaf' method, which selects the smallest, fine-grained clusters at the tips of the tree branches, representing clusters that cannot be split further. We derived both `EOM' and `leaf' clusters for each population to characterize the stellar complexes and substructures within them. Figures~\ref{fig:condensed_tree}~and~\ref{fig:age_cluster} show an example of our HDBSCAN run for the stellar age group  of 10-15 Myr, where the `EOM' and `leaf' clusters are highlighted (in colors) on the Condensed Tree and on the spatial distribution of the selections. More details of how the cluster selection are implemented for each method can be found in the HDBSCAN handbook \footnote{\url{https://hdbscan.readthedocs.io/en/latest/}}.

When using HDBSCAN, careful consideration of parameters is essential to customize the algorithm for specific needs. The two key parameters are minimum cluster size (\textit{min\_cluster\_size}) and minimum samples (\textit{min\_samples}). The parameter \textit{min\_cluster\_size} defines the smallest number of points (here stars) needed to form a cluster, with lower values creating smaller clusters and higher values forming larger ones. \textit{Min\_samples} determines the minimum number of points in a neighborhood for a point to be a core point, affecting cluster stability. Balancing these parameters is important to avoid small, noisy clusters or merging meaningful clusters into larger ones. On the other hand, the \textit{cluster\_selection\_epsilon} parameter helps manage dense areas by setting a distance threshold to merge closely situated clusters.

Key output attributes of HDBSCAN include labels\_, which assigns cluster labels to each data point (with noise points labeled as -1), condensed\_tree\_, which visualizes the cluster hierarchy (see Figure~\ref{fig:condensed_tree}), and probabilities\_, which gives the likelihood of each star belonging to a specific cluster, indicating the confidence level of the clustering results.

After performing various iterations using combinations of different values of \textit{min\_cluster\_size} and \textit{min\_samples} and through visual inspection of identified structures, we finally found the optimum value for \textit{min\_cluster\_size} and \textit{min\_samples} $=$ 5. Additionally, to avoid any subdivisions of the densest regions, we use \textit{cluster\_selection\_epsilon} of $\approx$20 pc. Finally, to achieve robust clustering of the field stars, we constrain the `probabilities\_' attributes to 0.68, which corresponds to a probability of 68$\%$ (or 1$\sigma$) that the star is part of the assigned cluster.

\begin{figure*}[htbp!]
\includegraphics[width=0.485\textwidth]{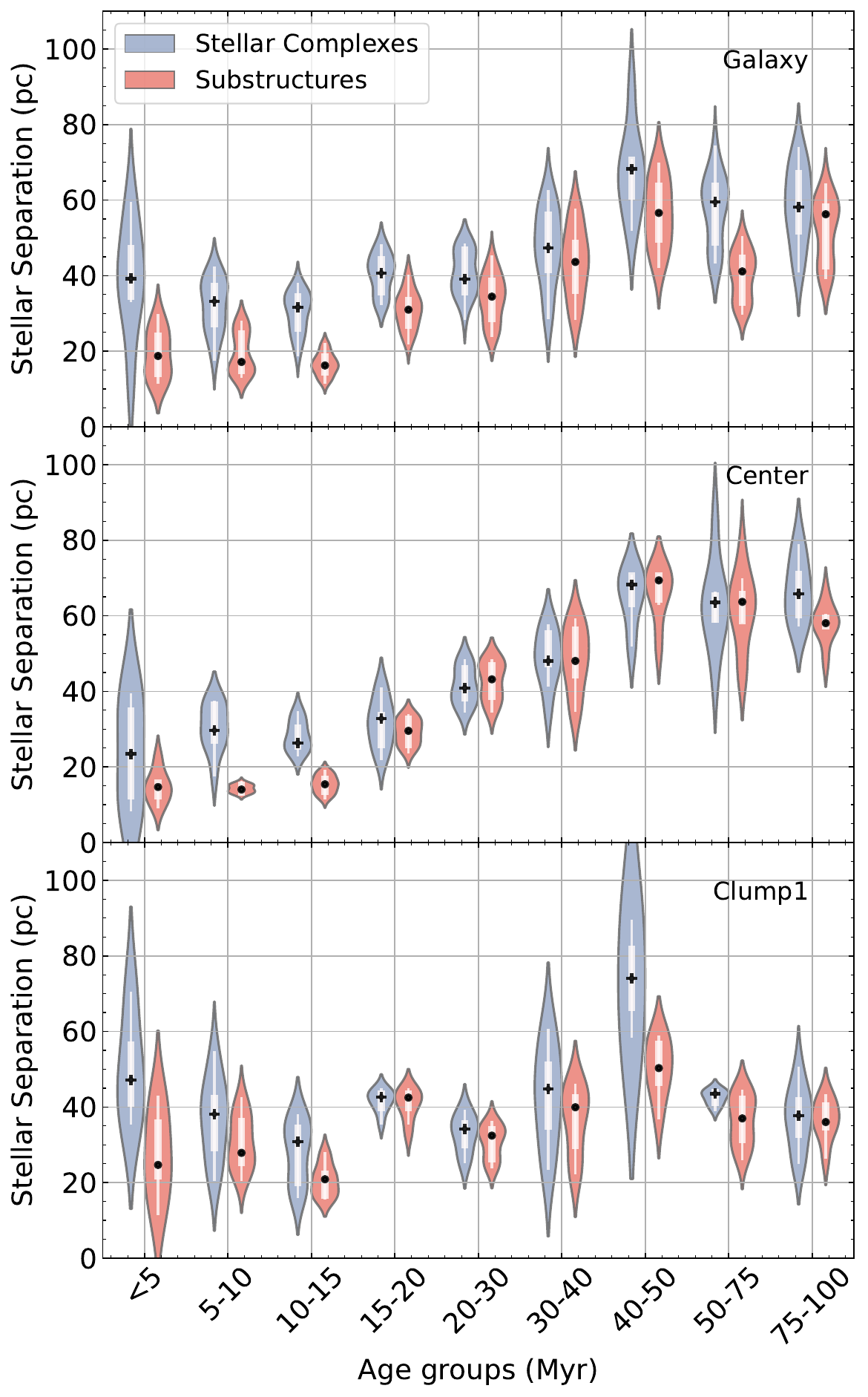}\hspace{1ex}
\includegraphics[width=0.485\textwidth]{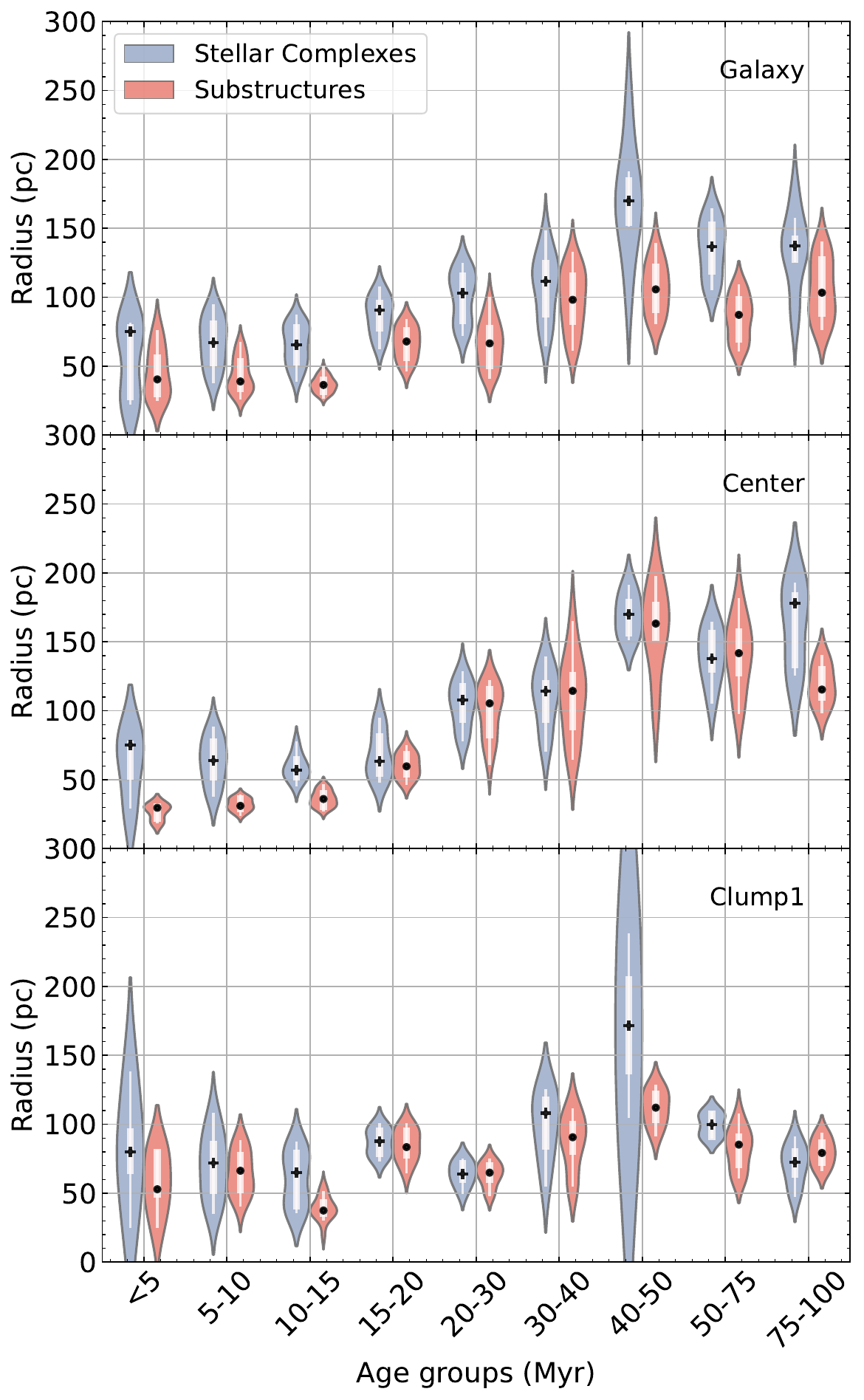}
\caption{A violin plot representation of the evolution of stellar structures in NGC~4449. The left panel shows the distribution of the average separation between stars and the right panel shows the distribution of the radii of the stellar complexes (shown in gray violin) and substructures (pink violin) in each age group (shown on the x-axis of the bottom most panel). The top panel of both panels shows the results for the whole galaxy (Center+Clump1) and the middle and bottom panels correspond separately to the Center and Clump1 regions respectively. The median value of the distribution in a particular age group is represented by black `+’ (for stellar complexes) and `.’ (for substructures). The thick white bar in the center of each violin represents the interquartile range and the thin white line represents the rest of the distribution. The thickness of the violin represent the boundary of the KDE, where the wider parts represent a higher probability of the variable (here, stellar separation) and the skinnier sections represent a lower probability.}
\label{fig:violin_all}
\end{figure*}

The identified stellar complex and their substructures are shown in Figure~\ref{fig:age_cluster} ~and~\ref{figapp:age_cluster1} \& \ref{figapp:age_cluster2}. 

Additionally, in order to gain deeper insights into the clustering properties of stars within different regions of the galaxy, we perform HDBSCAN clustering separately for the Center of the galaxy and for Clump1, as defined by \cite{Sacchi2018}, see also Figure~\ref{fig:NGC4449_regions}.

We use violin plots (see Figure~\ref{fig:violin_all}) to illustrate the distributions of average stellar separations and cluster radii for all the stellar complexes and substructures identified in different age bins. The shape and width of the plots indicate the frequency of observations, with wider sections representing higher data concentrations and narrower sections representing fewer data points. This comparison helps us find potential evolutionary trends and assess the influence of the galactic environment.

By examining the violin plots for the whole galaxy distribution (which encompasses both the Center and Clump1), we observe that the distribution of stellar separations and radii expands with age, signifying the expansion of stellar structures within the galaxy. However, this trend seems to reach a plateau after approximately 50 Myr. Although complexes and substructures remain present beyond this age, they exhibit characteristics similar to those identified using the TPCF (see \S\ref{subsec:tpcf}). Specifically, a few hierarchical structures persist, though the stellar distribution appears nearly random.

When we inspect the violin plots for different regions of the galaxy, we find that the Center closely mirrors the overall distribution observed for the entire galaxy. In contrast, Clump1 reveals no significant trends, suggesting that the expansion of stellar structures in NGC4449 is primarily driven by the Center region. The lack of correlation in Clump1 points to distinct dynamical processes or an alternative evolutionary history. These differences may arise from varying initial conditions, SFH, gravitational interactions, or external perturbations—potentially linked to tidal interactions with a nearby galaxy \citep{Hunter1998, Martinez-Delgado2012, Ai2023}—which may have influenced the arm-like structure of Clump1. For further discussion, see \S\ref{sec:discussion}.

\begin{figure}[htbp!]
\includegraphics[width=0.465\textwidth]{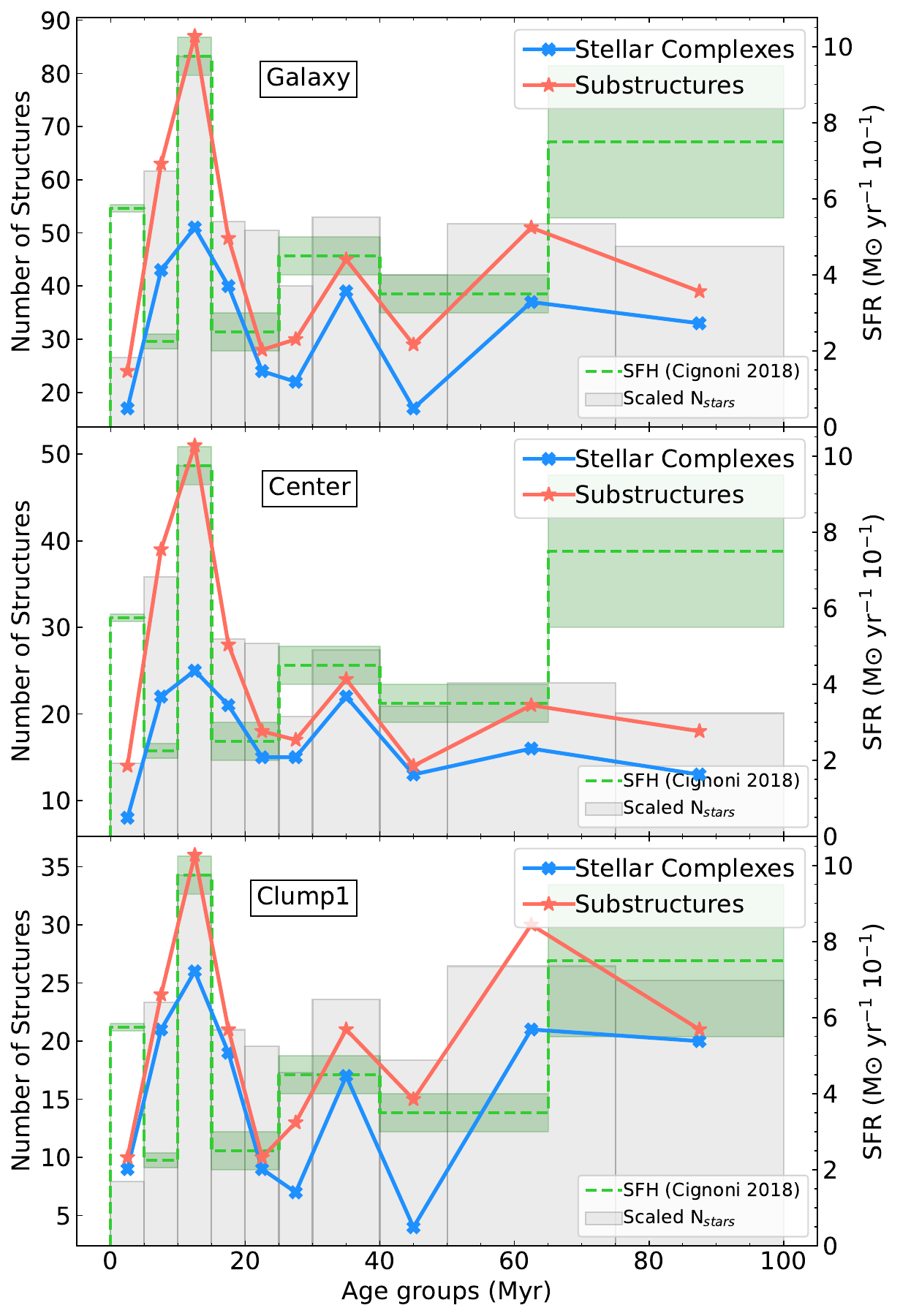}
\caption{A visual representation of the distribution of stellar ages versus the identified structures and SFH of NGC~4449 for the  entire galaxy (top panel), the Center region (middle panel) and the Clump1 (lower panel). The x axis shows the range of stellar age. The blue `X' (with solid line) corresponds to the number of stellar complexes and the red $\star$ (and the solid line) correspond to the number of substructures identified in different age groups as shown in Figure~\ref{fig:violin_all}. Their respective counts are marked on the left y-axis. The green step plot represents the recovered SFH of NGC~4449 from \cite{Cignoni2019}, with the SFR marked on the right y-axis. The corresponding uncertainties are shown with a green shaded area. Additionally, the gray bar plot shows the variation in the total number of stars, scaled to match the SFR for comparison, in each age bin. The overall agreement between the SFH and the evolution of stellar structures confirms that our clustering method effectively traces the recent SFH of NGC~4449.}
\label{fig:distributions}
\end{figure}

Figure~\ref{fig:distributions} shows the frequency of structures as a function of age. To assess if the drop in the number of structures after $\sim$~15 Myr is real and not an artifact due, for example, to the fast evolution of the most massive stars, or incompleteness effect in star counts, we compared the structure frequency with the recent SFH derived by \cite{Cignoni2019}. With the exception of the first bin, (0-5 Myr), which is biased by our selection criteria since we discard the stars for the BPASS age $=$ 0, the clustering frequency over the age range of 5 to 60 Myr is in very good agreement with the SFR reported by \cite{Cignoni2019}.

\subsection{Observed TPCF}\label{subsec:tpcf}

The TPCF (often denoted as \(\omega(\theta)\)), is one of the most widely used statistical tools to quantify the clustering behavior of a given spatial distribution of astronomical objects such as stars and galaxies (e.g. \citealp{Yung2023, Menon2021}). It assesses the excess probability of finding a pair of objects separated by an angular distance \(\theta\) relative to a random distribution. A positive \(\omega(\theta)\) at small separations signifies a higher probability of finding pairs of star close together, indicating strong clustering. In contrast, a \(\omega(\theta)\) close to zero suggests a random or homogeneous distribution. In this work, we perform TPCF on the field stars in NGC~4449 in order to assess their clustering behavior.

To calculate the TPCF, we utilize the code provided by \cite{Menon2021} on GitHub\footnote{\url{https://github.com/shm-1996/legus-tpcf/tree/main}}, which uses the astroML Python module \citep{astroML,astroMLText}, leveraging the scikit-learn library. We adapt and modify their code to calculate the TPCF for both star-clusters, following their original approach, and for field stars in this work. Since the random distribution in the original code is based on observations in the LEGUS FoV (using imaging observation in filter F435W), we recalculate it for the smaller field used in GULP (based on F275W mosaic). To ensure consistency with the analysis in \cite{Menon2021}, we used the same parameters for the galaxy properties like distance, inclination, and PA as adopted in \S\ref{subsec:deproject}.

A key factor in computing \(\omega(\theta)\) is the edge effect, where stars near the survey's boundary have fewer neighbors, potentially biasing the correlation function. This becomes significant at angular separations near the FoV boundaries, causing a steep drop in the TPCF as pairs diminish. To address this, we incorporate results from \cite{Menon2021}, who estimated the scale of edge effects \(\theta_{\text{edge}} \approx \theta_{\text{max}} / 5\), based on simulations of truncated 2D fractal distributions. Since our FoV is smaller than in \cite{Menon2021}, our \(\theta_{\text{edge}}\) is also reduced by the same factor.

\begin{figure*}[htbp!]
\includegraphics[width=0.49\textwidth]{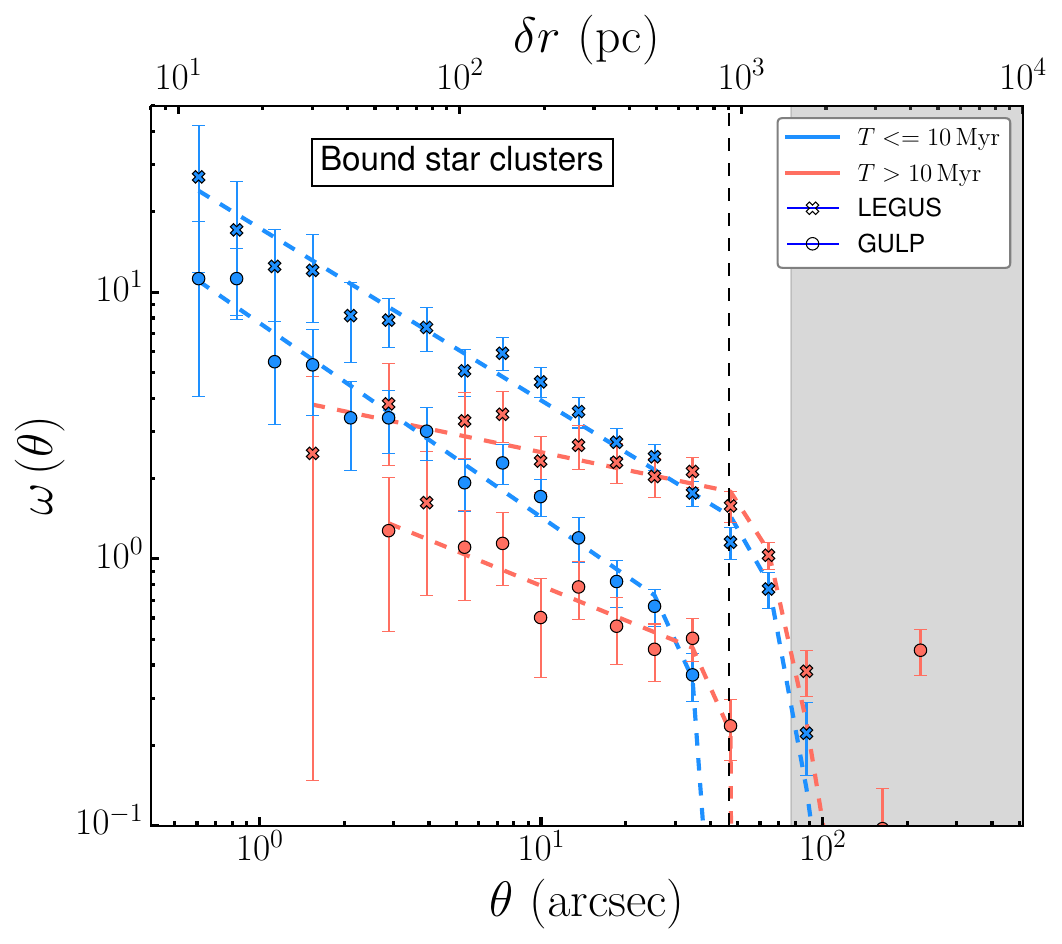}
\includegraphics[width=0.49\textwidth]{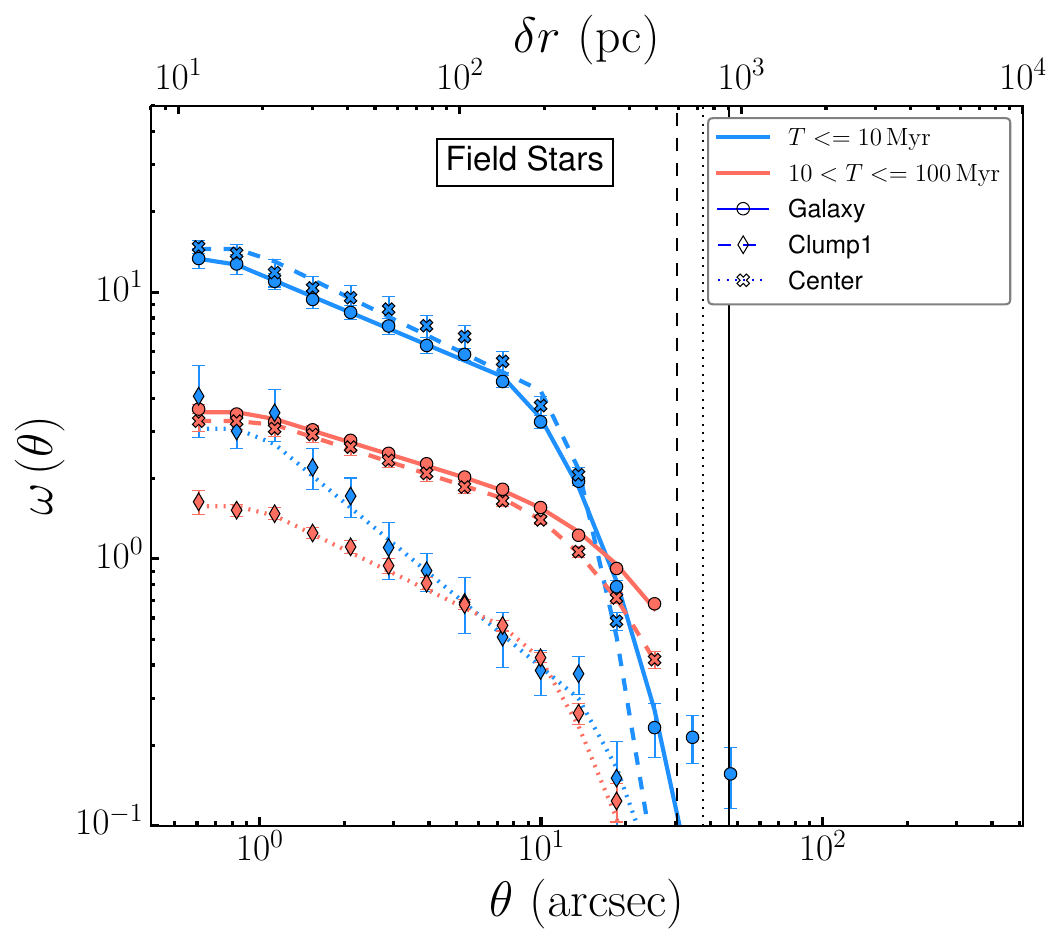}
\caption{Left: A comparison of the TPCF $\omega(\theta)$ for star-clusters in the LEGUS and GULP FoV. Star-clusters younger than 10 Myr are represented in blue, and those older than 10 Myr are in red. LEGUS TPCF data are indicated by 'x', while GULP data are marked by filled circles. MCMC assigns model 1 as best fits to both young and old star-clusters, though the slope for older star-clusters is notably shallower, approaching flatness. The best-fit curves for LEGUS are shown as solid lines, and for GULP as dashed lines. The star-cluster separation where edge effects become significant is highlighted by a gray-shaded region for LEGUS and a dashed line for GULP, with GULP’s boundary being smaller than that of LEGUS. Right: TPCF  of the field stars from GULP observations. Again the blue points represent young stars ($<$10 Myr), while the red points correspond to older stars ($>$10 Myr). As GULP primarily probes young massive stars, our analysis focuses on the stellar distribution within $<$100 Myr. The crossed, fitted with a solid line, illustrate the distribution for the entire galaxy (Center+Clump1), whereas the diamonds (dashed line fit) and circles (dotted line fit) represent the Clump1 and Center regions, respectively. The vertical black lines indicate the distances where edge effects become significant for each FoV: the solid line corresponds to the galaxy as a whole, the dashed line to Clump1, and the dotted line to the Center region.}
\label{fig:tpcf_1_10}
\end{figure*}

Figure~\ref{fig:tpcf_1_10} present the angular TPCF $\omega(\theta)$ as a function of angular separation $\theta$ for the young ($<=$ 10 Myr) and old ($>$ 10 Myr) star-clusters (in left panel) that were identified in \cite{Cook2023} and the field stars (on right panel) identified in this work. 
The most significant difference for the TPCF between star-clusters and field stars is the large scale transition point (breakpoint radius) where the power-law behavior shifts to an exponential decline. \cite{Menon2021} fit a second power-law to this decline. We find that an exponential decay more accurately represents the large-scale distribution, corresponding to the galaxy's exponential disk, and adopt this approach throughout the paper. For the star-clusters, this breakpoint happens at $\sim$60\arcsec~(See Table \ref{tab1:MCMC_params}), whereas for field stars it is notably smaller, at $\sim$10\arcsec. Interestingly, while the overall slope of the TPCF is similar for clusters and stars, the earlier breakpoint for field stars may indicate differences in their hierarchical structure or disruption timescales. Additionally, the TPCF for the old ($>$10 Myr) star-clusters flattens out, reflecting the loss of hierarchy in the older population, as noted in previous studies \citep{Grasha2017a, Menon2021}. In contrast, while the TPCF slope for field stars is shallower for the older population compared to the younger population, it is not entirely flat, indicating that some hierarchical structures persist at 100 Myr—particularly in the Clump1 region relative to the Center. 

Finally, we observe a turnover at small angular separations ($<$ 1\arcsec) in the field star population, a feature absent in the TPCF of star-clusters. This small-scale turnover in TPCF highlights the role of dispersal in the removal of hierarchy among young stars in smaller structures (as discussed in \citealp{Elmegreen2018}), a process we will explore further in \S\ref{sec:discussion}.

To  further investigate how hierarchical structures evolve with time, we analyze the TPCF over a finer temporal grid. We therefore divided the stars younger than 100 Myr in six age bins: $<$5 Myr, 5–10 Myr, 10–20 Myr, 20–30 Myr, 30–50 Myr, and 50–100 Myr, following the approach in \S\ref{subsec:steller_structures}. This stratification enables us to perform TPCF analysis within each age group and compare the results with those obtained from the clustering algorithm in \S\ref{subsec:steller_structures}.

\begin{figure}[htbp!]
\includegraphics[width=0.470\textwidth]{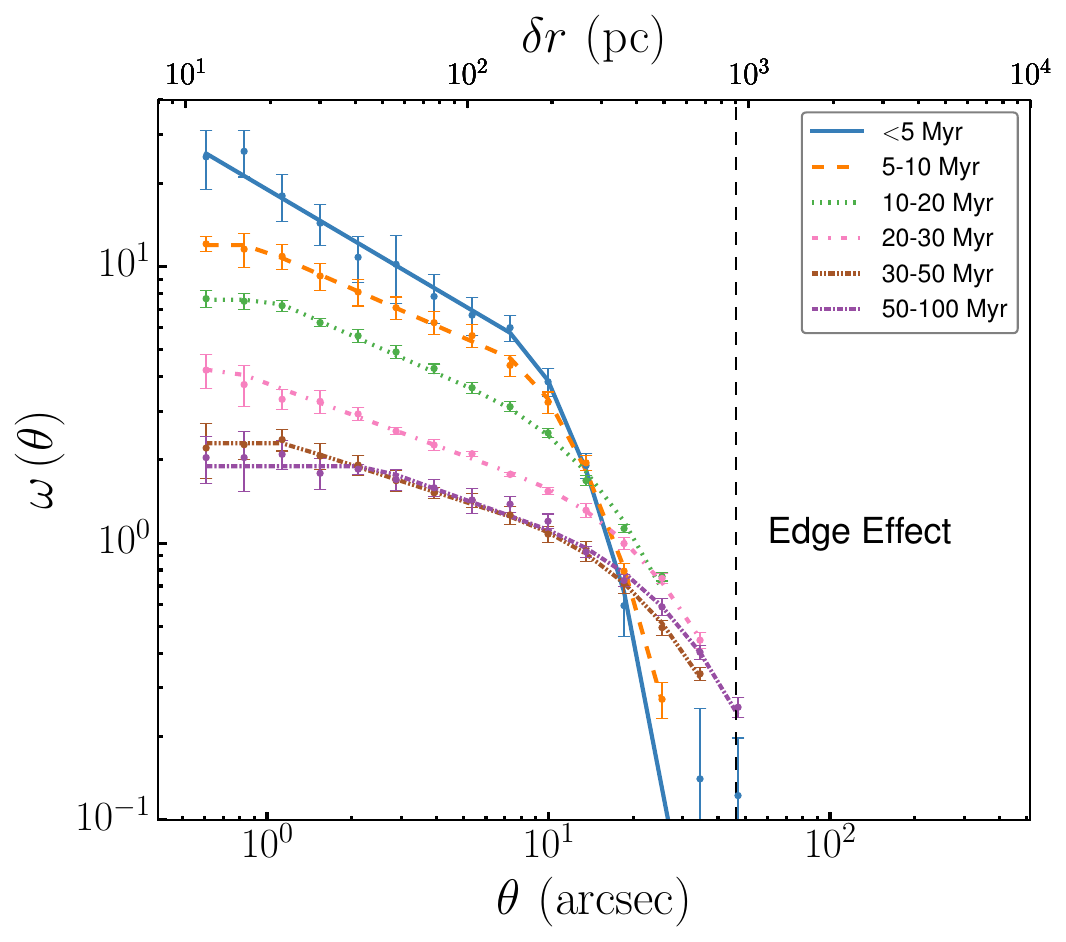}
\caption{TPCF $\omega(\theta)$ of the field stars that are younger than 100 Myr in NGC~4449, divided into different age groups i.e. $<$ 5 Myr, 5-10 Myr, 10-20 Myr, 20-30 Myr, 30-50 Myr and 50-100 Myr. 
The TPCF and associated uncertainties is represented by the error bars in different colors for different age groups. The best fits for $\omega(\theta)$ vs $\theta$ are shown in solid lines in same colors as the error bars of each age group in different line styles as shown in the legend. The bottom axis for $\theta$ shows the observed angular separation and top axis for $\delta$r shows the de-projected separation between stars. The vertical black dashed line shows the distance beyond which the edge effect for our observation's FOV becomes significant.}
\label{fig:tpcf_full}
\end{figure}

The angular TPCF $\omega(\theta)$ as a function of angular separation $\theta$ between stars in the same age group is shown in Figure~\ref{fig:tpcf_full}.
In general, our observed TPCF curves show that younger stars have a significantly higher $\omega(\theta)$ compared to older stars progressively, indicating a higher degree of correlation and clustering among the younger population. Furthermore, we observe a noticeable variation in the TPCF from smaller scales ($\sim$10 pc) to larger scales ($\sim$200 pc) within the same age group, with a marked difference as we transition from younger to older stars. The TPCF demonstrates a steep decline from small $\theta$ to large $\theta$ for younger stars and this decline becomes progressively shallower with increasing age, eventually `almost' flattening for stars older than 50 Myr.
This trend indicates a hierarchical distribution of stars in the field of the galaxy similar to what has been previously observed for the gravitational star-clusters in NGC~4449 (see Figure~\ref{fig:tpcf_full} (Left)) and some of the other local star-forming galaxies \citep{Gouliermis2015,Grasha2017a,Menon2021}. It is also consistent with our finding from clustering analysis in section \ref{subsec:steller_structures} that younger structures tend to be more compact (Figure~\ref{fig:violin_all}) and more numerous (Figure~\ref{fig:distributions}).

In addition to the change in the decline of the correlation strength as well as the exponent of the power law (seen as a linear slope between $\omega(\theta)$ and $\theta$ on the log-log plot in Figure~\ref{fig:tpcf_full}), we again notice a turnover or curl in the TPCF at small scales for all populations.

At angular separations greater than 200 pc, the TPCF curves for stars in 0-5 Myr and 5-10 Myr bins show a sharp decline. These populations are primarily concentrated in the galaxy's bar, with a few small star-forming regions in the northern arm (see Paper-I). 

As already highlighted by the HDBSCAN analysis (Figure \ref{figapp:age_cluster1}, stars younger than 20 Myr tend to be organized in more compact structures typically smaller than 100-150 pc (Figure \ref{fig:violin_all}). This is likely the cause of the observed  sharp decline in the TPCFs of the younger populations at $\theta > 200\, {\rm pc}$. 

\subsection{Quantifying TPCF}\label{subsec:mcmc}

In order to numerically quantify the variation of the TPCF and the effect of stellar ages, we fit three mathematical models to our observed TPCF curves, as below.

Model 1: A power law (representing hierarchical distribution) up to a specific distance, $\theta{_N}$, followed by an exponential fall off (`Model PF' in \citealp{Menon2021}):

\begin{equation}\label{eq:pl_exp}
w(\theta) = \begin{cases} A \cdot \theta^{-\alpha} &: \theta \leq \theta_N \\
A \cdot \theta_N^{-\alpha} \cdot \exp\left(-\beta (\theta - \theta_N)\right) &: \theta > \theta_N, \end{cases}
\end{equation}

where A is the amplitude of the TPCF, $\alpha$ is the power law index and $\beta$ is the index of the exponential decay for $\theta > \theta{_N}$.

Model 2: A constant value up to distance $\theta{_{N1}}$, indicating flattening in TPCF at small scales (equivalent to turnover at small scales), plus a power law up to a distance, $\theta{_{N2}}$, followed by an exponential fall off.

\begin{equation}\label{eq:pl_exp_const}
w(\theta) = \begin{cases} 
A ~~~~~~~~~~~~~~~~~~~~~~~~~~~~~~~: \theta \leq \theta_{N1} \\
A \cdot \left(\frac{\theta}{\theta_{N1}}\right)^{-\alpha} ~~~~~~~~~~~~~~~~~: \theta_{N1} < \theta \leq \theta_{N2} \\
A \cdot \left(\frac{\theta_{N2}}{\theta_{N1}}\right)^{-\alpha} \cdot \exp\left(-\beta (\theta - \theta_{N2})\right) : \theta > \theta_{N2} 
\end{cases}
\end{equation}

Model 3: A constant value up to a distance $\theta{_{N1}}$, followed by an exponential decline. This suggests no initial hierarchy, with the decay reflecting the exponential disk profile of the galaxy.

\begin{equation}\label{eq:const_exp}
w(\theta) =
\begin{cases} 
A & ~~~~~~~~~~~~~~~~~~~: \theta \leq \theta_N \\
A \cdot e^{-\beta (\theta - \theta_N)} & ~~~~~~~~~~~~~~~~~~~: \theta > \theta_N 
\end{cases}
\end{equation}

We use Markov Chain Monte Carlo (MCMC) method in the emcee Python package \citep{Foreman-Mackey2013} to estimate the best-fit parameters for each of the above models and the find the best fit to each TPCF curve.

\begin{deluxetable*} {c|c|c|c|c|c|c}[ht!]
\caption{Best fit model parameters: star-clusters and field stars (younger and older than 10 Myrs)}\label{tab1:MCMC_params}
\tablehead{\colhead{Age bins} &
\colhead{Best Model} & \colhead{A} & \colhead{$\theta_{N1}$ ($\arcsec$)} & \colhead{$\alpha$} & \colhead{$\theta_{N}$ or $\theta_{N2}$ ($\arcsec$)} & \colhead{$\beta$}}
\startdata
YSC$_{LEGUS}$ & 1 & 17.753 ± 3.159 & NA & 0.673 ± 0.105 & 60.904  ± 26.388 &  0.104 ± 2.073  \\
OSC$_{LEGUS}$ & 1 & 5.559 ± 1.718 & NA & 0.356 ±  0.170 & 61.223 ± 30.047 &  0.275 ±  2.152  \\
YSC$_{GULP}$ & 1 & 8.215 ± 1.193 & NA & 0.798 ± 0.085 & 44.098 ± 17.658 &   3.992   3.208  \\
OSC$_{GULP}$ & 1 &  3.460 ± 1.619 & NA & 0.635 ± 0.141 & 64.878 ± 18.699 & 4.905 ± 2.993   \\
YS$_{Galaxy}$ & 2 & 13.973 ± 5.843 & 0.655 ± 0.682 & 0.440 ± 0.115 & 7.662 ± 3.263 & 0.159 ± 2.886 \\
OS$_{Galaxy}$ & 2 & 8.290 ± 5.879 & 0.680 ± 0.774 & 0.324 ± 0.261 & 5.016 ± 2.306 & 0.058 ± 1.319 \\
YS$_{Clump1}$ & 2 & 4.821 ± 2.333 & 1.103 ± 0.406 &  0.932 ± 0.214 & 22.530 ± 15.929 &  3.425 ± 3.148 \\
OS$_{Clump1}$ & 2 & 2.404 ± 2.412 & 0.064 ± 0.636 & 0.462 ± 0.216 & 8.026 ± 1.768 & 0.149 ± 0.282 \\
YS$_{Center}$ & 2 & 14.802 ± 5.885 & 0.831 ± 0.388 &  0.488 ± 0.080 & 11.361 ± 1.056 & 0.290 ± 0.074 \\
OS$_{Center}$ & 2 & 3.244 ± 2.123 & 1.182 ± 0.598 & 0.382 ± 0.323 & 4.558 ± 4.297 & 0.077 ± 1.224 
\enddata
\tablecomments{YSC: Young stellar cluster ($<=10$ Myr) and OSC: old stellar cluster ($>10$ Myr); YS: young stars ($<=10$ Myr) and OS: old stars ($>10$ Myr)}
\end{deluxetable*}

The MCMC analysis identifies Model 1 (power law, exponential) as the best fit for the TPCF of star-clusters (Figure~\ref{fig:tpcf_1_10}, left). For older populations in LEGUS FoV, the power law exponent is so small that the TPCF curve nearly flattens out, while for GULP FoV, the exponent is slightly larger (i.e., steeper on a log-log plot), accompanied by larger uncertainties in the TPCF value for each age bin. This steeper slope in GULP is potentially influenced by the smaller FoV, which limits the ability to capture larger-scale distribution seen in LEGUS. The power law exponent for star-clusters (age $\leq$ 10 Myr) is consistent with those reported in \cite{Menon2021}, within uncertainties (see Table~\ref{tab1:MCMC_params}). Additionally, there is an offset in $\omega(\theta)$ in the overlapping regions of GULP and LEGUS. This discrepancy likely arises because GULP contains a smaller number of  star-clusters compared to LEGUS since LEGUS extends beyond the GULP FoV, which is limited to F275W footprint as shown in Figure~\ref{fig:NGC4449_regions}, includes more `clustered' star-clusters, leading to a higher overall $\omega(\theta)$ value. However, the overall shape of the TPCF remains the same across both datasets, which is the key result signifying the hierarchical pattern of this population. Lastly, the coefficient of the exponential decay beyond the break radii is much steeper for the GULP population compared to LEGUS, due to a sharper drop in the TPCF at separations approaching the GULP FoV. Please refer to Table~\ref{tab1:MCMC_params} for specific values of the fit parameters.

\begin{deluxetable*} {c|c|c|c|c|c|c}[htbp!]
\caption{Best fit model parameters: Different age bins for field stars}\label{tab2:MCMC_params}
\tablehead{\colhead{Age bins} &
\colhead{Best Model} & \colhead{A} & \colhead{$\theta_{N1}$ ($\arcsec$)} & \colhead{$\alpha$} & \colhead{$\theta_{N} ($\arcsec$)$ or $\theta_{N2}$} & \colhead{$\beta$}}
\startdata
5 & 2 & 33.20 ± 19.16 & 0.73 ± 0.57 & 0.65 ± 0.14 & 9.40 ± 3.71 & 1.15 ± 2.38 \\
10 & 2 & 12.67 ± 2.29 & 0.87 ± 0.37 & 0.45 ± 0.09 & 7.89 ± 1.82 & 0.17 ± 0.07 \\
20 & 2 & 7.90 ± 1.23 & 1.04 ± 0.28 & 0.47 ± 0.05 & 5.42 ± 1.48 & 0.09 ± 0.08 \\
30 & 2 & 7.48 ± 2.40 & 0.32 ± 0.36 & 0.39 ± 0.09 & 8.48 ± 2.58 & 0.06 ± 0.11 \\
50 & 2 & 3.70 ± 1.33 & 1.49 ± 0.29 & 0.41 ± 0.09 & 8.59 ± 1.67 & 0.07 ± 0.09 \\
100 & 2 & 2.36 ± 1.70 & 1.98 ± 0.73 & 0.44 ± 0.26 & 6.74 ± 3.40 & 0.14 ± 0.55 
\enddata
\end{deluxetable*} 

For field stars, Model 2 is preferred in all cases: (1) when divided into two populations (young and older than 10~Myr, Figure~\ref{fig:tpcf_1_10}, right), and (2) when divided into smaller age bins (Figure~\ref{fig:tpcf_full}). This model effectively captures the small-scale turnover observed in the field stars, which is absent in star-clusters, as discussed earlier (see \S\ref{subsec:tpcf}) and will be elaborated further in \S\ref{sec:discussion}. For the youngest age group ($<$ 5 Myr), as shown in Figure~\ref{fig:tpcf_full}, the turnover distance ($\theta_{N1}$) is so small that the TPCF visually resembles a power-law with an exponential decay, similar to Model 1.

\begin{figure*}[htbp!]
\includegraphics[width=\textwidth]{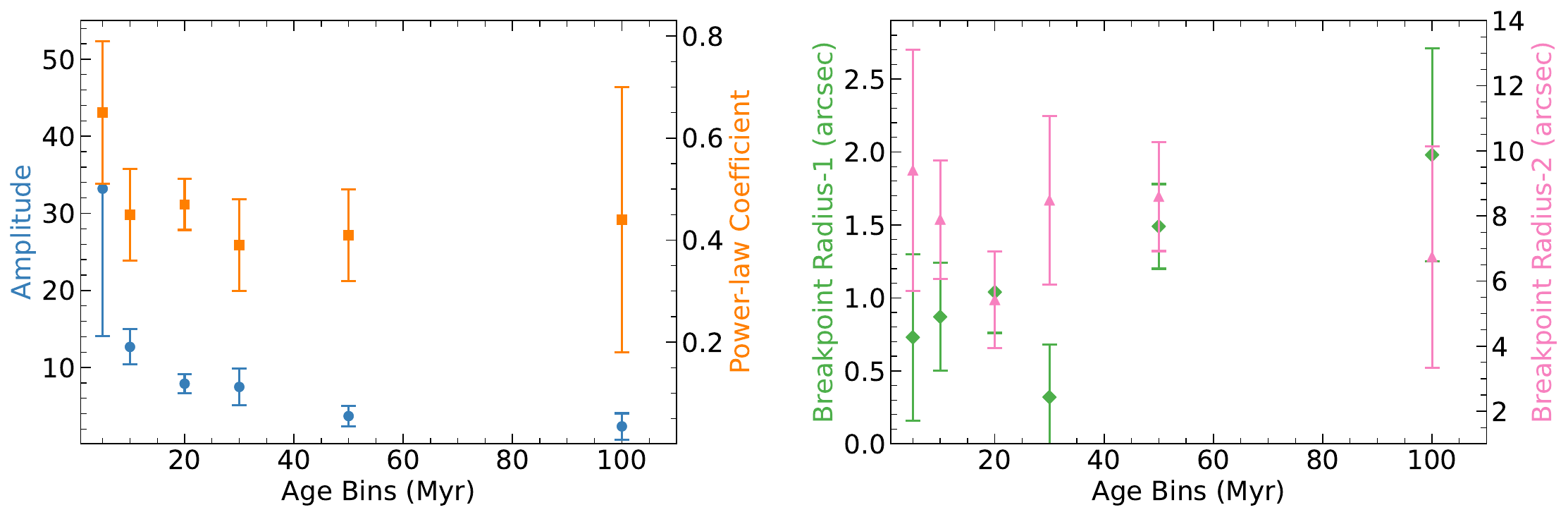}
\caption{TPCF fit parameters vs stellar age groups. The left panel illustrates the variation in the maximum clustering strength, `A' (blue dots), and the power-law coefficient of the TPCF, $\alpha$ (orange squares), across different stellar age groups. The right panel presents the evolution of the two breakpoint radii: $\theta_{N1}$ (green diamonds) and $\theta_{N2}$ (pink triangles), highlighting changes in the characteristic scales of stellar clustering over time.}
\label{fig:mcmc_params} 
\end{figure*}

Details of the MCMC runs are provided in Appendix~\ref{app:mcmc}. The best-fit models and associated parameters for the smaller age bins of the field stars are listed in Table~\ref{tab2:MCMC_params}. Figure~\ref{fig:mcmc_params} illustrates the variation of parameters A, $\alpha$, $\theta_{N1}$, and $\theta_{N2}$ with age. We find that the clustering amplitude $A$, and the power-law coefficient $\alpha$, decrease sharply from the youngest population ($<$5 Myr) to the 5–10 Myr age group. While $A$ continues to reduce for older populations, there is very little decline in $\alpha$ for older populations. This trend indicates: (1) the clustering strength of the young stars diminishes with time across all spatial scales (2) Although small, there are indications of a weakening in stellar hierarchy with age. However, the TPCF distribution does not become fully random for stars in our sample (i.e. $<$ 100 Myr).

Furthermore, with the exception for age group of 20-30 Myr, $\theta_{N1}$ shows a gradual increase with age, indicating the smaller structures in the same age group lose hierarchy quicker relative to the larger structure. We will discuss this more in details in \S\ref{sec:discussion}. While the MCMC analysis for the 20-30 Myr age group suggests a very small turnover in $\theta_{N1}$, it is important to note that the TPCF values within 2$\arcsec$ are actually comparable within the uncertainties. This suggests that, despite the small turnover indicated by the model, the underlying spatial distribution of stars in this age group still maintains a level of consistency in clustering at small scales. 

Finally, we observe a gradual decrease (within uncertainties) in the exponent ($\alpha$) of the power law part of curves (which is shown as a linear slope in the log-log plot). In younger populations, a steeper power-law slope points to stronger clustering at small scales. However, as stars age, the gradual decline in $\alpha$  reflects the loss of structural hierarchy over time.

The transition distance $\theta_{N2}$ represents the separation where the stellar hierarchy gives way to the galaxy’s disk profile. Across all age groups, $\theta_{N2}$ remains roughly consistent at around 10\arcsec (or $\sim$200 pc), with uncertainties factored in. However this distance is less than half that of observed for star-clusters therefore unlikely to be caused by the `edge effect' due to the GULP FoV. Similarly, the exponent $\beta$ shows consistent values across all age groups, indicating the galaxy's exponential disk profile is present in each case. However, for the youngest population ($<$5 Myr), $\beta$ deviates and has large uncertainties due to the relative scarcity of stars in that age range at larger spatial scale in the field, as previously discussed in \S\ref{subsec:tpcf}. We will explore the broader implications of our observed results and how they contribute to the our overall understanding of stellar hierarchy of NGC~4449 in the next section (\S\ref{sec:discussion}).

\section{Discussion}\label{sec:discussion}

Using HDBSCAN clustering and TPCF analysis, we find that stellar structures in NGC~4449 exhibit hierarchical and fractal characteristics, with most structures containing substructures, consistent with previous studies (e.g., \citealp{Gusev2014,Gouliermis2017,Rodriguez2019}). A clustering analysis across different stellar age groups reveals that complexes and substructures expand over time, reflecting the gradual dispersal of stars from their natal environments. Similarly, our TPCF analysis reveals how hierarchical behavior evolves with time. For younger stars ($<$ 20 Myr), the TPCF exhibits a strong power-law behavior, indicating significant grouping of field stars at smaller scales, which declines rapidly at larger separations. This suggests that young stars are located in dense, compact regions—likely remnants of their natal molecular clouds—even if they are no longer bound to their stellar nurseries.

This evolution likely occurs as stellar nurseries lose gravitational potential, leading to the progressive dispersal of stars until the structures eventually dissolve into the galactic field. The process appears to accelerate around 10–15 Myr, coinciding with the typical timescale for supernova explosions. The energy injected by these events likely contributes to the expansion and disruption of stellar structures. Notably, this period also corresponds to the peak star formation activity in NGC~4449 (Figure~\ref{fig:distributions}), suggesting a strong link between star formation episodes and the dynamical evolution of young stellar structures.

A characteristic spatial scale emerges in this process: the radii and stellar separations saturate around 200 pc, which is consistent with the typical size of a star complex composed of multiple molecular clouds \citep{Efremov1995}. This scale also corresponds to what has been described as a ``supercloud" in gas structures \citep{Elmegreen1987, Park2023}, suggesting a connection between the spatial extent of young stellar structures and the dimensions of their progenitor gas clouds. The 200 pc scale is further evident in the TPCF analysis of field stars (Figure~\ref{fig:tpcf_full}, right), where it marks a transition from a power-law to an exponential decline. This shift may reflect the influence of the galaxy’s exponential disk, reinforcing the idea that stars retain an imprint of their natal gas structures, which initially fragmented into smaller clumps before forming star-forming regions.

Furthermore, the number of stellar complexes and substructures increases with higher star formation activity, suggesting that intense star formation episodes lead to more structured stellar distributions and enhanced spatial sub-structuring in star-forming regions. Despite these correlations, the evolution of stellar separations and radii (Figure~\ref{fig:violin_all}) appears independent of star formation peaks, with them continuing to increase even after SF activity declines. This provides two key insights:
(1) Age Binning Validity: The selected age groups align well with the overall SFH, confirming that our age divisions capture physically meaningful evolutionary stages. (2) Clustering Robustness: The clustering analysis does not introduce any bias. If it had, we would expect the variation in stellar separations and radii (Figure~\ref{fig:violin_all}) to follow the trends seen in the SFH (Figure~\ref{fig:distributions}), rather than showing a consistent evolution with age.

Similar to our observations of expansion of stellar structures, the TPCF analysis provides additional evidence for this process. As stellar populations age, the gradual decrease in the TPCF power-law slope reinforces the notion of hierarchical evolution with age. The overall trend for the TPCF of young ($<$10 Myr) field stars resembles that of star-clusters in \cite{Menon2021}, with a power-law slope of approximately 0.6. However, two key discrepancies arise: (1) The large-scale transition point, where the power-law behavior shifts to an exponential decline, occurs at $\approx$60\arcsec~($\sim$1155 pc) for star-clusters, whereas for field stars, it is significantly smaller, around $\approx$10\arcsec ($\sim$192 pc). (2) \cite{Menon2021} found that structural hierarchy in star-clusters completely disappears (i.e., the TPCF becomes flat) for populations older than 10 Myr. In contrast, we observe that while the power-law slope decreases for older field stars, it does not become entirely flat, at least not for stellar populations younger than 100 Myr. There could be two reasons behind this. (1) Since for our age calculations, the sources were restricted to the ones observed in F275W, our TPCF investigation might be affected by incompleteness at the older end of the stellar population, which creates somewhat patchy distributions of the stars. (2) The structural hierarchy for the field stars indeed does not disappear by 100 Myr but takes longer time. However, since our observations are constrained to younger population, we do not probe those timescales when this hierarchy is completely destroyed. 

Beyond internal stellar dynamics, the broader galactic environment also influences the evolution of young stellar structures. External factors such as tidal interactions and large-scale gravitational perturbations can shape their distribution and dispersal over time. In this context, we examine the role of shear in the Clump1 and Center regions to assess its potential impact on the structural evolution of young stars.

\citet{Valdez-Gutierrez2002} (Fig. 13) provide an optical rotation curve for the inner $\pm$2 kpc of NGC~4449 based on H$\alpha$ observations. Given that our Center region spans the inner $\sim$2.5 kpc (Figure~\ref{fig:obs_deproject}), the differential rotation in this area is approximately 30 km s$^{-1}$. The apparent flattening toward the south is likely due to uncertainties in the velocity gradients in this region. Moreover, Clump1, located beyond $\sim$1 kpc to the northeast, experiences slightly stronger differential rotation of $\sim$40 km s$^{-1}$.

Using the first Oort parameter $A$ \citep{Oort1927, Seigar2005}, the shear rate is $3.5$ km s$^{-1}$ kpc$^{-1}$ for the Center region and around $10$ km s$^{-1}$ kpc$^{-1}$ for Clump1. The stronger shear in Clump1 could drive faster disruption of stellar structures, though it may not significantly affect structures on timescales of a few tens of Myr. 

Since shear is weak and negligible in the Center region, particularly on the timescales we probe, the evolution of stellar structures we observe is primarily driven by random stellar motions (Figure~\ref{fig:violin_all}). Conversely, Clump1 is more prone to external perturbations, including tidal interactions with nearby galaxies. However, even if shear were a dominant factor in Clump1, we would expect a flatter TPCF at larger angular separations (Figure~\ref{fig:tpcf_1_10}). As \citet{Elmegreen2018} noted, at small scales, stars maintain their correlation since stars within the same azimuthal circle move together. Any shear-induced disruption would primarily manifest at larger scales, yet we do not observe this flattening in Clump1, suggesting minimal impact of galactic rotation and shear on its observed stellar structure evolution.

Interestingly, substructuring patterns also differ between Clump1 and the Center. In the Center, substructuring peaks at 10–20 Myr, while Clump1 exhibits a bimodal distribution with peaks at 10–15 Myr and again at 40–60 Myr (Figure~\ref{fig:distributions}). The second peak coincides with a secondary increase in star formation activity \citep{Sacchi2018,Cignoni2019}, implying that strong star formation episodes enhance the formation of smaller substructures. Conversely, early star formation peaks (before 10 Myr) correlate with smoother stellar distributions over time, as young stars disperse.

Finally, we observe a curl or turnover in the TPCF at small scales, as discussed by \cite{Elmegreen2018}, who suggests that in the case of diffusion, smaller structures should disperse more quickly than larger structures due to the random motion of stars at their formation. This can be explained by considering that the initial diffusion velocity is comparable to a turbulent speed, or a virial speed in the cloud (here stars in structure group together). This velocity $v$ is proportional to the size of the structure L such that $v\propto L^{\alpha}$, where $\alpha$ is $\sim0.5$ for a typical ISM \citep{Solomon1987, Brunt2004}. Now the time taken to distort the structures (called as crossing time) will be $\frac{L}{v}$ or $L^{1-\alpha}$ or $L^{0.5}$. This implies that the larger structures take longer to disperse and the smaller structures smear out faster. 
In Figure \ref{fig:tpcf_full}, the TPCF curve for all age groups clearly curls over at small radii. The MCMC fits in Table \ref{tab2:MCMC_params} output these radii as $\theta_{N1}$. Except for the 20–30 Myr population, the turnover distance clearly increases with stellar age, roughly rising from $\sim$10 pc to 40 pc (see Figure~\ref{fig:mcmc_params}). Dividing the turnover distance by the stellar age gives a diffusion velocity of $\sim1$ km s$^{-1}$. This velocity is comparable to the velocity dispersion of a molecular cloud on this scale in the MW. That is, for an angular separation of $\sim20$ pc or a cloud with diameter of 20 pc, the relative velocity or velocity dispersion is $\sim1.6$ km s$^{-1}$ \citep{Rice2016}. Hence, our TPCF of the observed stars provides a clear indication of diffusion of stellar structures due to the random motion of stars.

While the physical processes discussed above provide insight into the early evolution of stellar structures, it is important to account for the observational limitations of our dataset. Although the full sample, including archival optical data, traces stellar populations up to $\sim$10 Gyr (as shown in \citealp{Sacchi2018} using optical datasets),this study focuses on stars younger than 100 Myr. This cutoff reflects the constraints of UV-based detection: beyond $\sim$100 Myr, increasing incompleteness introduces artificial spatial gaps, which can bias the interpretation of stellar clustering and evolution. This effect is particularly visible in the underdensity of stars beyond $\sim$60 pc in Figure~\ref{fig:violin_all} and \ref{figapp:age_cluster2}. By limiting our analysis to UV-detected sources, we ensure reliable mapping of young stellar structures and their evolution.

\section{Summary and Conclusion}\label{sec:summary}

We investigated the nature of stellar hierarchy and the processes by which stars migrate from their natal structures to populate the field of the dwarf starburst galaxy NGC~4449, using data from the GULP survey. We analyzed the spatial distribution of field stars younger than 100 Myr and identified stellar structures across different age groups using a density-based clustering algorithm. Furthermore, through TPCF analysis, we trace the evolution of stellar hierarchy over time, revealing how young massive stars disperse and how hierarchical structures evolve within the galaxy. Our main findings are summarized as below.

\begin{enumerate}
    \item Through HDBSCAN clustering, we map the hierarchical arrangement of young stars, where large-scale stellar complexes fragment into smaller substructures. Over time, these structures expand and dissolve, illustrating the progressive dispersal of young stars into the galactic field of NGC~4449.
    \item Our clustering analysis not only reveals the hierarchical nature of stellar structures but also effectively traces the SFH of NGC 4449. This suggests that similar techniques could be a powerful tool for reconstructing SFHs in other galaxies. Future studies should explore how well this method applies across different galactic environments.
    \item The TPCF analysis shows a decline in clustering strength with stellar age, further supporting the expansion and gradual dissolution of stellar structures over time.
    \item The transition distance at which hierarchical clustering fades ($\sim$200 pc) aligns with the expected size of supercloud complexes, indicating that young stars preserve a spatial imprint of their natal gas clouds.
    \item The expansion of stellar structures is primarily driven by random stellar motions and dispersion rather than shear from galaxy rotation.
    \item A comparison of the Center and Clump1 regions reveals distinct evolutionary patterns in the hierarchical distribution of stars:
    \begin{itemize}
        \item The Center exhibits a clear trend of hierarchical structure dispersal, with stellar complexes expanding over time and gradually dissolving into the field population.
        \item In contrast, Clump1 does not show a discernible trend of structural expansion or dispersal, suggesting a different evolutionary history.
        \item While Clump1 experiences relatively more differential rotation and shear effects compared to the Center, we do not observe any impact on the evolution and dispersal of stellar structures.
        \item The lack of an evolutionary trend in Clump1 may be attributed to external perturbations, possibly due to tidal interactions with a neighboring galaxy, which could have disrupted the hierarchical dispersal process.
    \end{itemize}
\end{enumerate}

In conclusion, this work suggest that stellar hierarchy in NGC~4449 evolves primarily through internal stellar motions, with young stars retaining an imprint of their natal gas structures before gradually dispersing into the field. However, limitations in our study, such as the focus on a single galaxy and potential observational incompleteness at older stellar ages, mean that a broader statistical approach is needed. Future studies will extend this analysis to the 26 additional galaxies in the GULP survey, allowing us to better constrain these evolutionary trends across different galactic environments and assess how factors such as galaxy morphology, SFH, and external interactions influence the hierarchical distribution and dispersal of young stars.

\begin{acknowledgments}

We are grateful to the anonymous referee for the careful review and constructive feedback that helped improve clarity of this work. 
The HST observations used in this paper are associated with program No. 16316.
This work is based on observations obtained with the NASA/ESA Hubble Space Telescope, at the Space Telescope Science Institute, which is operated by the Association of Universities for Research in Astronomy, Inc., under NASA contract NAS 5-26555. This work has made use of data from the European Space Agency (ESA) mission {\it Gaia} (\url{https://www.cosmos.esa.int/gaia}), processed by the {\it Gaia} Data Processing and Analysis Consortium (DPAC, \url{https://www.cosmos.esa.int/web/gaia/dpac/consortium}). Funding for the DPAC has been provided by national institutions, in particular, the institutions participating in the {\it Gaia} Multilateral Agreement.
E.S. is supported by the international Gemini Observatory, a program of NSF NOIRLab, which is managed by the Association of Universities for Research in Astronomy (AURA) under a cooperative agreement with the U.S. National Science Foundation, on behalf of the Gemini partnership of Argentina, Brazil, Canada, Chile, the Republic of Korea, and the United States of America.
L.B. contribution to this work was supported by grant HST-GO-16316.002-A.
A.A acknowledges support from the Swedish National Space Agency (SNSA) through the grant 2021- 00108.
RSK acknowledges financial support from the European Research Council via ERC Synergy Grant ``ECOGAL'' (project ID 855130),  from the German Excellence Strategy via the Heidelberg Cluster ``STRUCTURES'' (EXC 2181 - 390900948), and from the German Ministry for Economic Affairs and Climate Action in project ``MAINN'' (funding ID 50OO2206).  RSK also thanks the 2024/25 Class of Radcliffe Fellows for highly interesting and stimulating discussions. 
\end{acknowledgments}

\facilities{\textit{HST} (ACS, WFPC3)}

\software{SAOImage DS9 \citep{ds92000}, Python (\citealp{VanRossum2009}, \url{https://www.python.org}), Astropy \citep{astropy:2013, astropy:2018}, HDBSCAN \citep{Campello2013, McInnes2017} }

\bibliography{gulp_stellar_hierarchy}{}

\begin{thebibliography}{}
\expandafter\ifx\csname natexlab\endcsname\relax\def\natexlab#1{#1}\fi
\providecommand{\url}[1]{\href{#1}{#1}}
\providecommand{\dodoi}[1]{doi:~\href{http://doi.org/#1}{\nolinkurl{#1}}}
\providecommand{\doeprint}[1]{\href{http://ascl.net/#1}{\nolinkurl{http://ascl.net/#1}}}
\providecommand{\doarXiv}[1]{\href{https://arxiv.org/abs/#1}{\nolinkurl{https://arxiv.org/abs/#1}}}

\bibitem[{{Ai} {et~al.}(2023){Ai}, {Zhu}, {Xu}, {Wang}, {Jing}, {Yu}, \&
  {Jiang}}]{Ai2023}
{Ai}, M., {Zhu}, M., {Xu}, J.-l., {et~al.} 2023, \mnras, 524, 2911,
  \dodoi{10.1093/mnras/stad2011}

\bibitem[{{Astropy Collaboration} {et~al.}(2013){Astropy Collaboration},
  {Robitaille}, {Tollerud}, {Greenfield}, {Droettboom}, {Bray}, {Aldcroft},
  {Davis}, {Ginsburg}, {Price-Whelan}, {Kerzendorf}, {Conley}, {Crighton},
  {Barbary}, {Muna}, {Ferguson}, {Grollier}, {Parikh}, {Nair}, {Unther},
  {Deil}, {Woillez}, {Conseil}, {Kramer}, {Turner}, {Singer}, {Fox}, {Weaver},
  {Zabalza}, {Edwards}, {Azalee Bostroem}, {Burke}, {Casey}, {Crawford},
  {Dencheva}, {Ely}, {Jenness}, {Labrie}, {Lim}, {Pierfederici}, {Pontzen},
  {Ptak}, {Refsdal}, {Servillat}, \& {Streicher}}]{astropy:2013}
{Astropy Collaboration}, {Robitaille}, T.~P., {Tollerud}, E.~J., {et~al.} 2013,
  \aap, 558, A33, \dodoi{10.1051/0004-6361/201322068}

\bibitem[{{Bastian} {et~al.}(2007){Bastian}, {Ercolano}, {Gieles},
  {Rosolowsky}, {Scheepmaker}, {Gutermuth}, \& {Efremov}}]{Bastian2007}
{Bastian}, N., {Ercolano}, B., {Gieles}, M., {et~al.} 2007, \mnras, 379, 1302,
  \dodoi{10.1111/j.1365-2966.2007.12064.x}

\bibitem[{{Bastian} {et~al.}(2009){Bastian}, {Gieles}, {Ercolano}, \&
  {Gutermuth}}]{Bastian2009}
{Bastian}, N., {Gieles}, M., {Ercolano}, B., \& {Gutermuth}, R. 2009, \mnras,
  392, 868, \dodoi{10.1111/j.1365-2966.2008.14107.x}

\bibitem[{{Bastian} {et~al.}(2012){Bastian}, {Adamo}, {Gieles}, {Silva-Villa},
  {Lamers}, {Larsen}, {Smith}, {Konstantopoulos}, \&
  {Zackrisson}}]{bastian2012}
{Bastian}, N., {Adamo}, A., {Gieles}, M., {et~al.} 2012, \mnras, 419, 2606,
  \dodoi{10.1111/j.1365-2966.2011.19909.x}

\bibitem[{{Bianchi} {et~al.}(2014){Bianchi}, {Kang}, {Hodge}, {Dalcanton}, \&
  {Williams}}]{Bianchi2014}
{Bianchi}, L., {Kang}, Y., {Hodge}, P., {Dalcanton}, J., \& {Williams}, B.
  2014, Advances in Space Research, 53, 928, \dodoi{10.1016/j.asr.2013.08.024}

\bibitem[{{Bressan} {et~al.}(2012){Bressan}, {Marigo}, {Girardi}, {Salasnich},
  {Dal Cero}, {Rubele}, \& {Nanni}}]{Bressan2012}
{Bressan}, A., {Marigo}, P., {Girardi}, L., {et~al.} 2012, \mnras, 427, 127,
  \dodoi{10.1111/j.1365-2966.2012.21948.x}

\bibitem[{{Brunt} \& {Mac Low}(2004)}]{Brunt2004}
{Brunt}, C.~M., \& {Mac Low}, M.-M. 2004, \apj, 604, 196,
  \dodoi{10.1086/381648}

\bibitem[{{Calzetti} {et~al.}(2015){Calzetti}, {Lee}, {Sabbi}, {Adamo},
  {Smith}, {Andrews}, {Ubeda}, {Bright}, {Thilker}, {Aloisi}, {Brown},
  {Chandar}, {Christian}, {Cignoni}, {Clayton}, {da Silva}, {de Mink}, {Dobbs},
  {Elmegreen}, {Elmegreen}, {Evans}, {Fumagalli}, {Gallagher}, {Gouliermis},
  {Grebel}, {Herrero}, {Hunter}, {Johnson}, {Kennicutt}, {Kim}, {Krumholz},
  {Lennon}, {Levay}, {Martin}, {Nair}, {Nota}, {{\"O}stlin}, {Pellerin},
  {Prieto}, {Regan}, {Ryon}, {Schaerer}, {Schiminovich}, {Tosi}, {Van Dyk},
  {Walterbos}, {Whitmore}, \& {Wofford}}]{Calzetti2015}
{Calzetti}, D., {Lee}, J.~C., {Sabbi}, E., {et~al.} 2015, \aj, 149, 51,
  \dodoi{10.1088/0004-6256/149/2/51}

\bibitem[{{Calzetti} {et~al.}(2018){Calzetti}, {Wilson}, {Draine}, {Roussel},
  {Johnson}, {Heyer}, {Wall}, {Grasha}, {Battisti}, {Andrews}, {Kirkpatrick},
  {Rosa Gonz{\'a}lez}, {Vega}, {Puschnig}, {Yun}, {{\"O}stlin}, {Evans},
  {Tang}, {Lowenthal}, \& {S{\'a}nchez-Arguelles}}]{Calzetti2018}
{Calzetti}, D., {Wilson}, G.~W., {Draine}, B.~T., {et~al.} 2018, \apj, 852,
  106, \dodoi{10.3847/1538-4357/aaa1e2}

\bibitem[{Campello {et~al.}(2013)Campello, Moulavi, \& Sander}]{Campello2013}
Campello, R., Moulavi, D., \& Sander, J. 2013, in Advances in Knowledge
  Discovery and Data Mining. Pacific-Asia Conference on Knowledge Discovery and
  Data Mining, Vol. 7819 (Springer Berlin Heidelberg), 160--172,
  \dodoi{10.1007/978-3-642-37456-2_14}

\bibitem[{{Cartwright} \& {Whitworth}(2004)}]{Cartwright2004}
{Cartwright}, A., \& {Whitworth}, A.~P. 2004, \mnras, 348, 589,
  \dodoi{10.1111/j.1365-2966.2004.07360.x}

\bibitem[{{Chandar} {et~al.}(2006){Chandar}, {Fall}, \&
  {Whitmore}}]{Chandar2006}
{Chandar}, R., {Fall}, S.~M., \& {Whitmore}, B.~C. 2006, \apjl, 650, L111,
  \dodoi{10.1086/508890}

\bibitem[{{Chi} {et~al.}(2023){Chi}, {Wang}, {Wang}, {Deng}, \& {Li}}]{Chi2023}
{Chi}, H., {Wang}, F., {Wang}, W., {Deng}, H., \& {Li}, Z. 2023, \apjs, 266,
  36, \dodoi{10.3847/1538-4365/accb50}

\bibitem[{{Cignoni} {et~al.}(2019){Cignoni}, {Sacchi}, {Tosi}, {Aloisi},
  {Cook}, {Calzetti}, {Lee}, {Sabbi}, {Thilker}, {Adamo}, {Dale}, {Elmegreen},
  {Gallagher}, {Grebel}, {Johnson}, {Messa}, {Smith}, \& {Ubeda}}]{Cignoni2019}
{Cignoni}, M., {Sacchi}, E., {Tosi}, M., {et~al.} 2019, \apj, 887, 112,
  \dodoi{10.3847/1538-4357/ab53d5}

\bibitem[{Comaniciu \& Meer(2002)}]{comaniciu_mean_2002}
Comaniciu, D., \& Meer, P. 2002, IEEE Transactions on Pattern Analysis and
  Machine Intelligence, 24, 603, \dodoi{10.1109/34.1000236}

\bibitem[{{Cook} {et~al.}(2019){Cook}, {Lee}, {Adamo}, {Kim}, {Chandar},
  {Whitmore}, {Mok}, {Ryon}, {Dale}, {Calzetti}, {Andrews}, {Aloisi},
  {Ashworth}, {Bright}, {Brown}, {Christian}, {Cignoni}, {Clayton}, {da Silva},
  {de Mink}, {Dobbs}, {Elmegreen}, {Elmegreen}, {Evans}, {Fumagalli},
  {Gallagher}, {Gouliermis}, {Grasha}, {Grebel}, {Herrero}, {Hunter}, {Jensen},
  {Johnson}, {Kahre}, {Kennicutt}, {Krumholz}, {Lee}, {Lennon}, {Linden},
  {Martin}, {Messa}, {Nair}, {Nota}, {{\"O}stlin}, {Parziale}, {Pellerin},
  {Regan}, {Sabbi}, {Sacchi}, {Schaerer}, {Schiminovich}, {Shabani}, {Slane},
  {Small}, {Smith}, {Smith}, {Taibi}, {Thilker}, {de la Torre}, {Tosi},
  {Turner}, {Ubeda}, {Van Dyk}, {Walterbos}, \& {Wofford}}]{Cook2019}
{Cook}, D.~O., {Lee}, J.~C., {Adamo}, A., {et~al.} 2019, \mnras, 484, 4897,
  \dodoi{10.1093/mnras/stz331}

\bibitem[{{Cook} {et~al.}(2023){Cook}, {Lee}, {Adamo}, {Calzetti}, {Chandar},
  {Whitmore}, {Aloisi}, {Cignoni}, {Dale}, {Elmegreen}, {Fumagalli}, {Grasha},
  {Johnson}, {Kennicutt}, {Kim}, {Linden}, {Messa}, {{\"O}stlin}, {Ryon},
  {Sacchi}, {Thilker}, {Tosi}, \& {Wofford}}]{Cook2023}
---. 2023, \mnras, 519, 3749, \dodoi{10.1093/mnras/stac3748}

\bibitem[{{Drazinos} {et~al.}(2013){Drazinos}, {Kontizas}, {Karampelas},
  {Kontizas}, \& {Dapergolas}}]{Drazinos2013}
{Drazinos}, P., {Kontizas}, E., {Karampelas}, A., {Kontizas}, M., \&
  {Dapergolas}, A. 2013, \aap, 553, A87, \dodoi{10.1051/0004-6361/201220648}

\bibitem[{{Efremov}(1995)}]{Efremov1995}
{Efremov}, Y.~N. 1995, \aj, 110, 2757, \dodoi{10.1086/117728}

\bibitem[{{Efremov} \& {Elmegreen}(1998)}]{Efremov1998}
{Efremov}, Y.~N., \& {Elmegreen}, B.~G. 1998, \mnras, 299, 588,
  \dodoi{10.1046/j.1365-8711.1998.01819.x}

\bibitem[{{Eldridge} \& {Rela{\~n}o}(2011)}]{2011MNRAS.411..235E}
{Eldridge}, J.~J., \& {Rela{\~n}o}, M. 2011, \mnras, 411, 235,
  \dodoi{10.1111/j.1365-2966.2010.17676.x}

\bibitem[{{Eldridge} {et~al.}(2017){Eldridge}, {Stanway}, {Xiao}, {McClelland},
  {Taylor}, {Ng}, {Greis}, \& {Bray}}]{Eldridge2017}
{Eldridge}, J.~J., {Stanway}, E.~R., {Xiao}, L., {et~al.} 2017, \pasa, 34,
  e058, \dodoi{10.1017/pasa.2017.51}

\bibitem[{{Elmegreen}(2008)}]{Elmegreen2008}
{Elmegreen}, B.~G. 2008, \apj, 672, 1006, \dodoi{10.1086/523791}

\bibitem[{{Elmegreen}(2018)}]{Elmegreen2018}
---. 2018, \apj, 853, 88, \dodoi{10.3847/1538-4357/aaa252}

\bibitem[{{Elmegreen} \& {Elmegreen}(1987)}]{Elmegreen1987}
{Elmegreen}, B.~G., \& {Elmegreen}, D.~M. 1987, \apj, 320, 182,
  \dodoi{10.1086/165534}

\bibitem[{{Elmegreen} \& {Scalo}(2004)}]{Elmegreen2004}
{Elmegreen}, B.~G., \& {Scalo}, J. 2004, \araa, 42, 211,
  \dodoi{10.1146/annurev.astro.41.011802.094859}

\bibitem[{{Fall} {et~al.}(2005){Fall}, {Chandar}, \& {Whitmore}}]{Fall2005}
{Fall}, S.~M., {Chandar}, R., \& {Whitmore}, B.~C. 2005, \apjl, 631, L133,
  \dodoi{10.1086/496878}

\bibitem[{{Foreman-Mackey} {et~al.}(2013){Foreman-Mackey}, {Hogg}, {Lang}, \&
  {Goodman}}]{Foreman-Mackey2013}
{Foreman-Mackey}, D., {Hogg}, D.~W., {Lang}, D., \& {Goodman}, J. 2013, \pasp,
  125, 306, \dodoi{10.1086/670067}

\bibitem[{{Gieles} {et~al.}(2008){Gieles}, {Bastian}, \&
  {Ercolano}}]{Gieles2008}
{Gieles}, M., {Bastian}, N., \& {Ercolano}, B. 2008, \mnras, 391, L93,
  \dodoi{10.1111/j.1745-3933.2008.00563.x}

\bibitem[{{Girichidis} {et~al.}(2020){Girichidis}, {Offner}, {Kritsuk},
  {Klessen}, {Hennebelle}, {Kruijssen}, {Krause}, {Glover}, \&
  {Padovani}}]{Girichidis2020}
{Girichidis}, P., {Offner}, S. S.~R., {Kritsuk}, A.~G., {et~al.} 2020, \ssr,
  216, 68, \dodoi{10.1007/s11214-020-00693-8}

\bibitem[{{Gordon} {et~al.}(2016){Gordon}, {Fouesneau}, {Arab}, {Tchernyshyov},
  {Weisz}, {Dalcanton}, {Williams}, {Bell}, {Bianchi}, {Boyer}, {Choi},
  {Dolphin}, {Girardi}, {Hogg}, {Kalirai}, {Kapala}, {Lewis}, {Rix},
  {Sandstrom}, \& {Skillman}}]{Gordon2016}
{Gordon}, K.~D., {Fouesneau}, M., {Arab}, H., {et~al.} 2016, \apj, 826, 104,
  \dodoi{10.3847/0004-637X/826/2/104}

\bibitem[{{Gouliermis}(2018)}]{Gouliermis2018}
{Gouliermis}, D.~A. 2018, \pasp, 130, 072001, \dodoi{10.1088/1538-3873/aac1fd}

\bibitem[{{Gouliermis} {et~al.}(2014){Gouliermis}, {Hony}, \&
  {Klessen}}]{Gouliermis2014}
{Gouliermis}, D.~A., {Hony}, S., \& {Klessen}, R.~S. 2014, \mnras, 439, 3775,
  \dodoi{10.1093/mnras/stu228}

\bibitem[{{Gouliermis} {et~al.}(2010){Gouliermis}, {Schmeja}, {Klessen}, {de
  Blok}, \& {Walter}}]{Gouliermis2010}
{Gouliermis}, D.~A., {Schmeja}, S., {Klessen}, R.~S., {de Blok}, W.~J.~G., \&
  {Walter}, F. 2010, \apj, 725, 1717, \dodoi{10.1088/0004-637X/725/2/1717}

\bibitem[{{Gouliermis} {et~al.}(2015){Gouliermis}, {Thilker}, {Elmegreen},
  {Elmegreen}, {Calzetti}, {Lee}, {Adamo}, {Aloisi}, {Cignoni}, {Cook}, {Dale},
  {Gallagher}, {Grasha}, {Grebel}, {Dav{\'o}}, {Hunter}, {Johnson}, {Kim},
  {Nair}, {Nota}, {Pellerin}, {Ryon}, {Sabbi}, {Sacchi}, {Smith}, {Tosi},
  {Ubeda}, \& {Whitmore}}]{Gouliermis2015}
{Gouliermis}, D.~A., {Thilker}, D., {Elmegreen}, B.~G., {et~al.} 2015, \mnras,
  452, 3508, \dodoi{10.1093/mnras/stv1325}

\bibitem[{{Gouliermis} {et~al.}(2017){Gouliermis}, {Elmegreen}, {Elmegreen},
  {Calzetti}, {Cignoni}, {Gallagher}, {Kennicutt}, {Klessen}, {Sabbi},
  {Thilker}, {Ubeda}, {Aloisi}, {Adamo}, {Cook}, {Dale}, {Grasha}, {Grebel},
  {Johnson}, {Sacchi}, {Shabani}, {Smith}, \& {Wofford}}]{Gouliermis2017}
{Gouliermis}, D.~A., {Elmegreen}, B.~G., {Elmegreen}, D.~M., {et~al.} 2017,
  \mnras, 468, 509, \dodoi{10.1093/mnras/stx445}

\bibitem[{{Grasha} {et~al.}(2015){Grasha}, {Calzetti}, {Adamo}, {Kim},
  {Elmegreen}, {Gouliermis}, {Aloisi}, {Bright}, {Christian}, {Cignoni},
  {Dale}, {Dobbs}, {Elmegreen}, {Fumagalli}, {Gallagher}, {Grebel}, {Johnson},
  {Lee}, {Messa}, {Smith}, {Ryon}, {Thilker}, {Ubeda}, \&
  {Wofford}}]{Grasha2015}
{Grasha}, K., {Calzetti}, D., {Adamo}, A., {et~al.} 2015, \apj, 815, 93,
  \dodoi{10.1088/0004-637X/815/2/93}

\bibitem[{{Grasha} {et~al.}(2017{\natexlab{a}}){Grasha}, {Calzetti}, {Adamo},
  {Kim}, {Elmegreen}, {Gouliermis}, {Dale}, {Fumagalli}, {Grebel}, {Johnson},
  {Kahre}, {Kennicutt}, {Messa}, {Pellerin}, {Ryon}, {Smith}, {Shabani},
  {Thilker}, \& {Ubeda}}]{Grasha2017a}
---. 2017{\natexlab{a}}, \apj, 840, 113, \dodoi{10.3847/1538-4357/aa6f15}

\bibitem[{{Grasha} {et~al.}(2017{\natexlab{b}}){Grasha}, {Elmegreen},
  {Calzetti}, {Adamo}, {Aloisi}, {Bright}, {Cook}, {Dale}, {Fumagalli},
  {Gallagher}, {Gouliermis}, {Grebel}, {Kahre}, {Kim}, {Krumholz}, {Lee},
  {Messa}, {Ryon}, \& {Ubeda}}]{Grasha2017b}
{Grasha}, K., {Elmegreen}, B.~G., {Calzetti}, D., {et~al.} 2017{\natexlab{b}},
  \apj, 842, 25, \dodoi{10.3847/1538-4357/aa740b}

\bibitem[{{Gusev}(2014)}]{Gusev2014}
{Gusev}, A.~S. 2014, \mnras, 442, 3711, \dodoi{10.1093/mnras/stu1095}

\bibitem[{{Hill} {et~al.}(1998){Hill}, {Fanelli}, {Smith}, {Bohlin}, {Neff},
  {O'Connell}, {Roberts}, {Smith}, \& {Stecher}}]{Hill1998}
{Hill}, R.~S., {Fanelli}, M.~N., {Smith}, D.~A., {et~al.} 1998, \apj, 507, 179,
  \dodoi{10.1086/306302}

\bibitem[{{Hunter} {et~al.}(2005){Hunter}, {Rubin}, {Swaters}, {Sparke}, \&
  {Levine}}]{Hunter2005}
{Hunter}, D.~A., {Rubin}, V.~C., {Swaters}, R.~A., {Sparke}, L.~S., \&
  {Levine}, S.~E. 2005, \apj, 634, 281, \dodoi{10.1086/496949}

\bibitem[{{Hunter} {et~al.}(1998){Hunter}, {Wilcots}, {van Woerden},
  {Gallagher}, \& {Kohle}}]{Hunter1998}
{Hunter}, D.~A., {Wilcots}, E.~M., {van Woerden}, H., {Gallagher}, J.~S., \&
  {Kohle}, S. 1998, \apjl, 495, L47, \dodoi{10.1086/311213}

\bibitem[{{Ivezi{\'c}} {et~al.}(2014){Ivezi{\'c}}, {Connolly}, {Vanderplas}, \&
  {Gray}}]{astroMLText}
{Ivezi{\'c}}, {\v Z}., {Connolly}, A., {Vanderplas}, J., \& {Gray}, A. 2014,
  Statistics, Data Mining and Machine Learning in Astronomy (Princeton
  University Press)

\bibitem[{Jin \& Han(2010)}]{Jin2010}
Jin, X., \& Han, J. 2010, in Encyclopedia of machine learning, ed. C.~Sammut \&
  G.~I. Webb (Boston, MA: Springer US), 563--564,
  \dodoi{10.1007/978-0-387-30164-8_425}

\bibitem[{{Klessen} \& {Glover}(2016)}]{Klessen2016}
{Klessen}, R.~S., \& {Glover}, S. C.~O. 2016, Saas-Fee Advanced Course, 43, 85,
  \dodoi{10.1007/978-3-662-47890-5_2}

\bibitem[{{Kroupa}(2001)}]{2001MNRAS.322..231K}
{Kroupa}, P. 2001, \mnras, 322, 231, \dodoi{10.1046/j.1365-8711.2001.04022.x}

\bibitem[{{Krumholz} {et~al.}(2019){Krumholz}, {McKee}, \&
  {Bland-Hawthorn}}]{Krumholz2019}
{Krumholz}, M.~R., {McKee}, C.~F., \& {Bland-Hawthorn}, J. 2019, \araa, 57,
  227, \dodoi{10.1146/annurev-astro-091918-104430}

\bibitem[{{Ksoll} {et~al.}(2021){Ksoll}, {Gouliermis}, {Sabbi}, {Ryon},
  {Robberto}, {Gennaro}, {Klessen}, {Koethe}, {de Marchi}, {Chen}, {Cignoni},
  \& {Dolphin}}]{Ksoll2021}
{Ksoll}, V.~F., {Gouliermis}, D., {Sabbi}, E., {et~al.} 2021, \aj, 161, 257,
  \dodoi{10.3847/1538-3881/abee8c}

\bibitem[{{Lada} \& {Lada}(2003)}]{Lada2003}
{Lada}, C.~J., \& {Lada}, E.~A. 2003, \araa, 41, 57,
  \dodoi{10.1146/annurev.astro.41.011802.094844}

\bibitem[{{Larson} {et~al.}(2023){Larson}, {Lee}, {Thilker}, {Whitmore},
  {Deger}, {Lilly}, {Chandar}, {Dale}, {Bigiel}, {Grasha}, {Groves}, {Hannon},
  {Klessen}, {Kreckel}, {Kruijssen}, {Leroy}, {Pan}, {Rosolowsky},
  {Schinnerer}, {Schruba}, {Watkins}, \& {Williams}}]{Larson2023}
{Larson}, K.~L., {Lee}, J.~C., {Thilker}, D.~A., {et~al.} 2023, \mnras, 523,
  6061, \dodoi{10.1093/mnras/stad1600}

\bibitem[{{Lee} {et~al.}(2009){Lee}, {Kennicutt}, {Funes}, {Sakai}, \&
  {Akiyama}}]{Lee2009}
{Lee}, J.~C., {Kennicutt}, Robert~C., J., {Funes}, S.~J. J.~G., {Sakai}, S., \&
  {Akiyama}, S. 2009, \apj, 692, 1305, \dodoi{10.1088/0004-637X/692/2/1305}

\bibitem[{{Lelli} {et~al.}(2014){Lelli}, {Verheijen}, \&
  {Fraternali}}]{Lelli2014}
{Lelli}, F., {Verheijen}, M., \& {Fraternali}, F. 2014, \mnras, 445, 1694,
  \dodoi{10.1093/mnras/stu1804}

\bibitem[{{Linden} {et~al.}(2022){Linden}, {Perez}, {Calzetti}, {Maji},
  {Messa}, {Whitmore}, {Chandar}, {Adamo}, {Grasha}, {Cook}, {Elmegreen},
  {Dale}, {Sacchi}, {Sabbi}, {Grebel}, \& {Smith}}]{linden2022}
{Linden}, S.~T., {Perez}, G., {Calzetti}, D., {et~al.} 2022, \apj, 935, 166,
  \dodoi{10.3847/1538-4357/ac7c07}

\bibitem[{{Mac Low} \& {Klessen}(2004)}]{MacLow2004}
{Mac Low}, M.-M., \& {Klessen}, R.~S. 2004, Reviews of Modern Physics, 76, 125,
  \dodoi{10.1103/RevModPhys.76.125}

\bibitem[{{Mart{\'\i}nez-Delgado} {et~al.}(2012){Mart{\'\i}nez-Delgado},
  {Romanowsky}, {Gabany}, {Annibali}, {Arnold}, {Fliri}, {Zibetti}, {van der
  Marel}, {Rix}, {Chonis}, {Carballo-Bello}, {Aloisi}, {Macci{\`o}},
  {Gallego-Laborda}, {Brodie}, \& {Merrifield}}]{Martinez-Delgado2012}
{Mart{\'\i}nez-Delgado}, D., {Romanowsky}, A.~J., {Gabany}, R.~J., {et~al.}
  2012, \apjl, 748, L24, \dodoi{10.1088/2041-8205/748/2/L24}

\bibitem[{McInnes {et~al.}(2017)McInnes, Healy, \& Astels}]{McInnes2017}
McInnes, L., Healy, J., \& Astels, S. 2017, The Journal of Open Source
  Software, 2, \dodoi{10.21105/joss.00205}

\bibitem[{{McKee} \& {Ostriker}(2007)}]{McKee2007}
{McKee}, C.~F., \& {Ostriker}, E.~C. 2007, \araa, 45, 565,
  \dodoi{10.1146/annurev.astro.45.051806.110602}

\bibitem[{{McQuinn} {et~al.}(2012){McQuinn}, {Skillman}, {Dalcanton}, {Cannon},
  {Dolphin}, {Holtzman}, {Weisz}, \& {Williams}}]{McQuinn2012}
{McQuinn}, K. B.~W., {Skillman}, E.~D., {Dalcanton}, J.~J., {et~al.} 2012,
  \apj, 759, 77, \dodoi{10.1088/0004-637X/759/1/77}

\bibitem[{{McQuinn} {et~al.}(2010){McQuinn}, {Skillman}, {Cannon}, {Dalcanton},
  {Dolphin}, {Hidalgo-Rodr{\'\i}guez}, {Holtzman}, {Stark}, {Weisz}, \&
  {Williams}}]{McQuinn2010}
{McQuinn}, K. B.~W., {Skillman}, E.~D., {Cannon}, J.~M., {et~al.} 2010, \apj,
  721, 297, \dodoi{10.1088/0004-637X/721/1/297}

\bibitem[{{Menon} {et~al.}(2021){Menon}, {Grasha}, {Elmegreen}, {Federrath},
  {Krumholz}, {Calzetti}, {S{\'a}nchez}, {Linden}, {Adamo}, {Messa}, {Cook},
  {Dale}, {Grebel}, {Fumagalli}, {Sabbi}, {Johnson}, {Smith}, \&
  {Kennicutt}}]{Menon2021}
{Menon}, S.~H., {Grasha}, K., {Elmegreen}, B.~G., {et~al.} 2021, \mnras, 507,
  5542, \dodoi{10.1093/mnras/stab2413}

\bibitem[{{Messa} {et~al.}(2018){Messa}, {Adamo}, {Calzetti}, {Reina-Campos},
  {Colombo}, {Schinnerer}, {Chandar}, {Dale}, {Gouliermis}, {Grasha}, {Grebel},
  {Elmegreen}, {Fumagalli}, {Johnson}, {Kruijssen}, {{\"O}stlin}, {Shabani},
  {Smith}, \& {Whitmore}}]{messa2018}
{Messa}, M., {Adamo}, A., {Calzetti}, D., {et~al.} 2018, \mnras, 477, 1683,
  \dodoi{10.1093/mnras/sty577}

\bibitem[{{Moe} \& {Di Stefano}(2017)}]{2017ApJS..230...15M}
{Moe}, M., \& {Di Stefano}, R. 2017, \apjs, 230, 15,
  \dodoi{10.3847/1538-4365/aa6fb6}

\bibitem[{{Oort}(1927)}]{Oort1927}
{Oort}, J.~H. 1927, \bain, 3, 275

\bibitem[{{Park} {et~al.}(2023){Park}, {Lee}, {Bialy}, {Burkhart}, {Dawson},
  {Heiles}, {Li}, {Murray}, {Nguyen}, {Hafner}, {Rybarczyk}, \&
  {Stanimirovi{\'c}}}]{Park2023}
{Park}, G., {Lee}, M.-Y., {Bialy}, S., {et~al.} 2023, \apj, 955, 145,
  \dodoi{10.3847/1538-4357/ace164}

\bibitem[{{Parker} {et~al.}(2014){Parker}, {Wright}, {Goodwin}, \&
  {Meyer}}]{Parker2014}
{Parker}, R.~J., {Wright}, N.~J., {Goodwin}, S.~P., \& {Meyer}, M.~R. 2014,
  \mnras, 438, 620, \dodoi{10.1093/mnras/stt2231}

\bibitem[{{Pastorelli} {et~al.}(2019){Pastorelli}, {Marigo}, {Girardi}, {Chen},
  {Rubele}, {Trabucchi}, {Aringer}, {Bladh}, {Bressan}, {Montalb{\'a}n},
  {Boyer}, {Dalcanton}, {Eriksson}, {Groenewegen}, {H{\"o}fner}, {Lebzelter},
  {Nanni}, {Rosenfield}, {Wood}, \& {Cioni}}]{Pastorelli2019}
{Pastorelli}, G., {Marigo}, P., {Girardi}, L., {et~al.} 2019, \mnras, 485,
  5666, \dodoi{10.1093/mnras/stz725}

\bibitem[{{Pastorelli} {et~al.}(2020){Pastorelli}, {Marigo}, {Girardi},
  {Aringer}, {Chen}, {Rubele}, {Trabucchi}, {Bladh}, {Boyer}, {Bressan},
  {Dalcanton}, {Groenewegen}, {Lebzelter}, {Mowlavi}, {Chubb}, {Cioni}, {de
  Grijs}, {Ivanov}, {Nanni}, {van Loon}, \& {Zaggia}}]{Pastorelli2020}
---. 2020, \mnras, 498, 3283, \dodoi{10.1093/mnras/staa2565}

\bibitem[{{Pellerin} {et~al.}(2007){Pellerin}, {Meyer}, {Harris}, \&
  {Calzetti}}]{Pellerin2007}
{Pellerin}, A., {Meyer}, M., {Harris}, J., \& {Calzetti}, D. 2007, \apjl, 658,
  L87, \dodoi{10.1086/515437}

\bibitem[{{Pellerin} {et~al.}(2012){Pellerin}, {Meyer}, {Calzetti}, \&
  {Harris}}]{Pellerin2012}
{Pellerin}, A., {Meyer}, M.~M., {Calzetti}, D., \& {Harris}, J. 2012, \aj, 144,
  182, \dodoi{10.1088/0004-6256/144/6/182}

\bibitem[{{Price-Whelan} {et~al.}(2018){Price-Whelan}, {Sip{\H{o}}cz},
  {G{\"u}nther}, {Lim}, {Crawford}, {Conseil}, {Shupe}, {Craig}, {Dencheva},
  {Ginsburg}, {VanderPlas}, {Bradley}, {P{\'e}rez-Su{\'a}rez}, {de Val-Borro},
  {Paper Contributors}, {Aldcroft}, {Cruz}, {Robitaille}, {Tollerud},
  {Coordination Committee}, {Ardelean}, {Babej}, {Bach}, {Bachetti}, {Bakanov},
  {Bamford}, {Barentsen}, {Barmby}, {Baumbach}, {Berry}, {Biscani}, {Boquien},
  {Bostroem}, {Bouma}, {Brammer}, {Bray}, {Breytenbach}, {Buddelmeijer},
  {Burke}, {Calderone}, {Cano Rodr{\'\i}guez}, {Cara}, {Cardoso}, {Cheedella},
  {Copin}, {Corrales}, {Crichton}, {D{\textquoteright}Avella}, {Deil},
  {Depagne}, {Dietrich}, {Donath}, {Droettboom}, {Earl}, {Erben}, {Fabbro},
  {Ferreira}, {Finethy}, {Fox}, {Garrison}, {Gibbons}, {Goldstein}, {Gommers},
  {Greco}, {Greenfield}, {Groener}, {Grollier}, {Hagen}, {Hirst}, {Homeier},
  {Horton}, {Hosseinzadeh}, {Hu}, {Hunkeler}, {Ivezi{\'c}}, {Jain}, {Jenness},
  {Kanarek}, {Kendrew}, {Kern}, {Kerzendorf}, {Khvalko}, {King}, {Kirkby},
  {Kulkarni}, {Kumar}, {Lee}, {Lenz}, {Littlefair}, {Ma}, {Macleod},
  {Mastropietro}, {McCully}, {Montagnac}, {Morris}, {Mueller}, {Mumford},
  {Muna}, {Murphy}, {Nelson}, {Nguyen}, {Ninan}, {N{\"o}the}, {Ogaz}, {Oh},
  {Parejko}, {Parley}, {Pascual}, {Patil}, {Patil}, {Plunkett}, {Prochaska},
  {Rastogi}, {Reddy Janga}, {Sabater}, {Sakurikar}, {Seifert}, {Sherbert},
  {Sherwood-Taylor}, {Shih}, {Sick}, {Silbiger}, {Singanamalla}, {Singer},
  {Sladen}, {Sooley}, {Sornarajah}, {Streicher}, {Teuben}, {Thomas},
  {Tremblay}, {Turner}, {Terr{\'o}n}, {van Kerkwijk}, {de la Vega}, {Watkins},
  {Weaver}, {Whitmore}, {Woillez}, {Zabalza}, \& {Contributors}}]{astropy:2018}
{Price-Whelan}, A.~M., {Sip{\H{o}}cz}, B.~M., {G{\"u}nther}, H.~M., {et~al.}
  2018, \aj, 156, 123, \dodoi{10.3847/1538-3881/aabc4f}

\bibitem[{{Rice} {et~al.}(2016){Rice}, {Goodman}, {Bergin}, {Beaumont}, \&
  {Dame}}]{Rice2016}
{Rice}, T.~S., {Goodman}, A.~A., {Bergin}, E.~A., {Beaumont}, C., \& {Dame},
  T.~M. 2016, \apj, 822, 52, \dodoi{10.3847/0004-637X/822/1/52}

\bibitem[{{Rich} {et~al.}(2012){Rich}, {Collins}, {Black}, {Longstaff}, {Koch},
  {Benson}, \& {Reitzel}}]{Rich2012}
{Rich}, R.~M., {Collins}, M.~L.~M., {Black}, C.~M., {et~al.} 2012, \nat, 482,
  192, \dodoi{10.1038/nature10837}

\bibitem[{{Rodr{\'\i}guez} {et~al.}(2016){Rodr{\'\i}guez}, {Baume}, \&
  {Feinstein}}]{Rodriguez2016}
{Rodr{\'\i}guez}, M.~J., {Baume}, G., \& {Feinstein}, C. 2016, \aap, 594, A34,
  \dodoi{10.1051/0004-6361/201527876}

\bibitem[{{Rodr{\'\i}guez} {et~al.}(2019){Rodr{\'\i}guez}, {Baume}, \&
  {Feinstein}}]{Rodriguez2019}
---. 2019, \aap, 626, A35, \dodoi{10.1051/0004-6361/201935291}

\bibitem[{{Rodr{\'\i}guez} {et~al.}(2020){Rodr{\'\i}guez}, {Baume}, \&
  {Feinstein}}]{Rodriguez2020}
---. 2020, \aap, 644, A101, \dodoi{10.1051/0004-6361/202038970}

\bibitem[{{Sabbi} {et~al.}(2018){Sabbi}, {Calzetti}, {Ubeda}, {Adamo},
  {Cignoni}, {Thilker}, {Aloisi}, {Elmegreen}, {Elmegreen}, {Gouliermis},
  {Grebel}, {Messa}, {Smith}, {Tosi}, {Dolphin}, {Andrews}, {Ashworth},
  {Bright}, {Brown}, {Chandar}, {Christian}, {Clayton}, {Cook}, {Dale}, {de
  Mink}, {Dobbs}, {Evans}, {Fumagalli}, {Gallagher}, {Grasha}, {Herrero},
  {Hunter}, {Johnson}, {Kahre}, {Kennicutt}, {Kim}, {Krumholz}, {Lee},
  {Lennon}, {Martin}, {Nair}, {Nota}, {{\"O}stlin}, {Pellerin}, {Prieto},
  {Regan}, {Ryon}, {Sacchi}, {Schaerer}, {Schiminovich}, {Shabani}, {Van Dyk},
  {Walterbos}, {Whitmore}, \& {Wofford}}]{Sabbi2018}
{Sabbi}, E., {Calzetti}, D., {Ubeda}, L., {et~al.} 2018, \apjs, 235, 23,
  \dodoi{10.3847/1538-4365/aaa8e5}

\bibitem[{{Sacchi} {et~al.}(2018){Sacchi}, {Cignoni}, {Aloisi}, {Tosi},
  {Calzetti}, {Lee}, {Adamo}, {Annibali}, {Dale}, {Elmegreen}, {Gouliermis},
  {Grasha}, {Grebel}, {Hunter}, {Sabbi}, {Smith}, {Thilker}, {Ubeda}, \&
  {Whitmore}}]{Sacchi2018}
{Sacchi}, E., {Cignoni}, M., {Aloisi}, A., {et~al.} 2018, \apj, 857, 63,
  \dodoi{10.3847/1538-4357/aab844}

\bibitem[{{Scalo} \& {Elmegreen}(2004)}]{Scalo2004}
{Scalo}, J., \& {Elmegreen}, B.~G. 2004, \araa, 42, 275,
  \dodoi{10.1146/annurev.astro.42.120403.143327}

\bibitem[{{Scheepmaker} {et~al.}(2009){Scheepmaker}, {Lamers}, {Anders}, \&
  {Larsen}}]{Scheepmaker2009}
{Scheepmaker}, R.~A., {Lamers}, H.~J.~G.~L.~M., {Anders}, P., \& {Larsen},
  S.~S. 2009, \aap, 494, 81, \dodoi{10.1051/0004-6361:200811068}

\bibitem[{{Schmeja} \& {Klessen}(2006)}]{Schmeja2006}
{Schmeja}, S., \& {Klessen}, R.~S. 2006, \aap, 449, 151,
  \dodoi{10.1051/0004-6361:20054464}

\bibitem[{{Seigar}(2005)}]{Seigar2005}
{Seigar}, M.~S. 2005, \mnras, 361, L20,
  \dodoi{10.1111/j.1745-3933.2005.00056.x}

\bibitem[{{Shashank} {et~al.}(2025){Shashank}, {Subramanian}, {Muraleedharan},
  {Menon}, {Mondal}, {Krishna}, {Das}, \& {Subramaniam}}]{Shashank2025}
{Shashank}, G., {Subramanian}, S., {Muraleedharan}, S., {et~al.} 2025, \aap,
  693, A188, \dodoi{10.1051/0004-6361/202451739}

\bibitem[{{Smithsonian Astrophysical Observatory}(2000)}]{ds92000}
{Smithsonian Astrophysical Observatory}. 2000, {SAOImage DS9: A utility for
  displaying astronomical images in the X11 window environment}.
\newblock \doeprint{0003.002}

\bibitem[{{Solomon} {et~al.}(1987){Solomon}, {Rivolo}, {Barrett}, \&
  {Yahil}}]{Solomon1987}
{Solomon}, P.~M., {Rivolo}, A.~R., {Barrett}, J., \& {Yahil}, A. 1987, \apj,
  319, 730, \dodoi{10.1086/165493}

\bibitem[{{Stanway} \& {Eldridge}(2018)}]{Stanway2018}
{Stanway}, E.~R., \& {Eldridge}, J.~J. 2018, \mnras, 479, 75,
  \dodoi{10.1093/mnras/sty1353}

\bibitem[{{Theis} \& {Kohle}(2001)}]{Theis2000}
{Theis}, C., \& {Kohle}, S. 2001, \aap, 370, 365,
  \dodoi{10.1051/0004-6361:20010198}

\bibitem[{{Valdez-Guti{\'e}rrez} {et~al.}(2002){Valdez-Guti{\'e}rrez},
  {Rosado}, {Puerari}, {Georgiev}, {Borissova}, \&
  {Ambrocio-Cruz}}]{Valdez-Gutierrez2002}
{Valdez-Guti{\'e}rrez}, M., {Rosado}, M., {Puerari}, I., {et~al.} 2002, \aj,
  124, 3157, \dodoi{10.1086/344304}

\bibitem[{Van~Rossum \& Drake(2009)}]{VanRossum2009}
Van~Rossum, G., \& Drake, F.~L. 2009, Python 3 Reference Manual (Scotts Valley,
  CA: CreateSpace)

\bibitem[{{Vanderplas} {et~al.}(2012){Vanderplas}, {Connolly}, {Ivezi{\'c}}, \&
  {Gray}}]{astroML}
{Vanderplas}, J., {Connolly}, A., {Ivezi{\'c}}, {\v Z}., \& {Gray}, A. 2012, in
  Conference on Intelligent Data Understanding (CIDU), 47 --54,
  \dodoi{10.1109/CIDU.2012.6382200}

\bibitem[{{Vargas-Salazar} {et~al.}(2020){Vargas-Salazar}, {Oey}, {Barnes},
  {Chen}, {Castro}, {Kratter}, \& {Faerber}}]{VargasSalazar2020}
{Vargas-Salazar}, I., {Oey}, M.~S., {Barnes}, J.~R., {et~al.} 2020, \apj, 903,
  42, \dodoi{10.3847/1538-4357/abbb95}

\bibitem[{{V{\'a}zquez-Semadeni} {et~al.}(2017){V{\'a}zquez-Semadeni},
  {Gonz{\'a}lez-Samaniego}, \& {Col{\'\i}n}}]{Vazquez-Semadeni2017}
{V{\'a}zquez-Semadeni}, E., {Gonz{\'a}lez-Samaniego}, A., \& {Col{\'\i}n}, P.
  2017, \mnras, 467, 1313, \dodoi{10.1093/mnras/stw3229}

\bibitem[{{Whitaker} {et~al.}(2012){Whitaker}, {van Dokkum}, {Brammer}, \&
  {Franx}}]{Whitaker2012}
{Whitaker}, K.~E., {van Dokkum}, P.~G., {Brammer}, G., \& {Franx}, M. 2012,
  \apjl, 754, L29, \dodoi{10.1088/2041-8205/754/2/L29}

\bibitem[{{Yung} {et~al.}(2023){Yung}, {Somerville}, {Finkelstein}, {Behroozi},
  {Dav{\'e}}, {Ferguson}, {Gardner}, {Popping}, {Malhotra}, {Papovich},
  {Rhoads}, {Bagley}, {Hirschmann}, \& {Koekemoer}}]{Yung2023}
{Yung}, L.~Y.~A., {Somerville}, R.~S., {Finkelstein}, S.~L., {et~al.} 2023,
  \mnras, 519, 1578, \dodoi{10.1093/mnras/stac3595}

\bibitem[{{Zhang} {et~al.}(2001){Zhang}, {Fall}, \& {Whitmore}}]{Zhang2001}
{Zhang}, Q., {Fall}, S.~M., \& {Whitmore}, B.~C. 2001, \apj, 561, 727,
  \dodoi{10.1086/322278}

\end{thebibliography}
\bibliographystyle{aasjournal}

\appendix
\restartappendixnumbering

\section{Stellar Structures in Different Age Groups} \label{app:structures}

\begin{figure*}[ht!]
\includegraphics[width=0.95\textwidth]{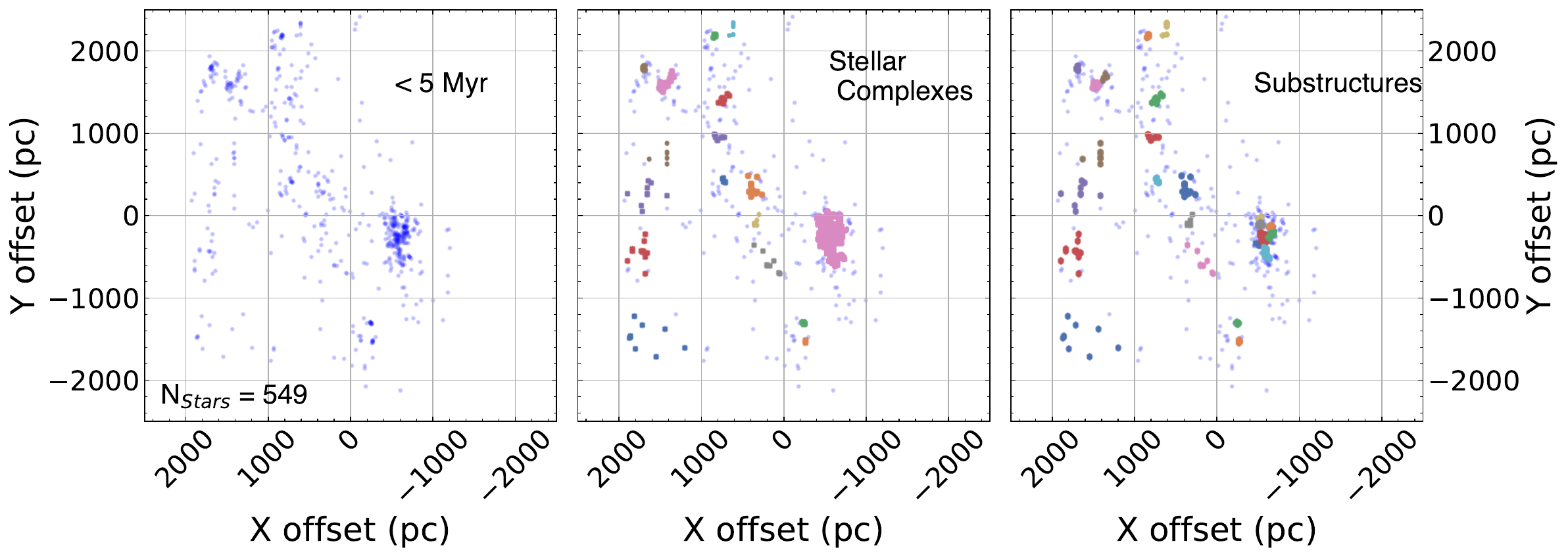}
\includegraphics[width=0.95\textwidth]{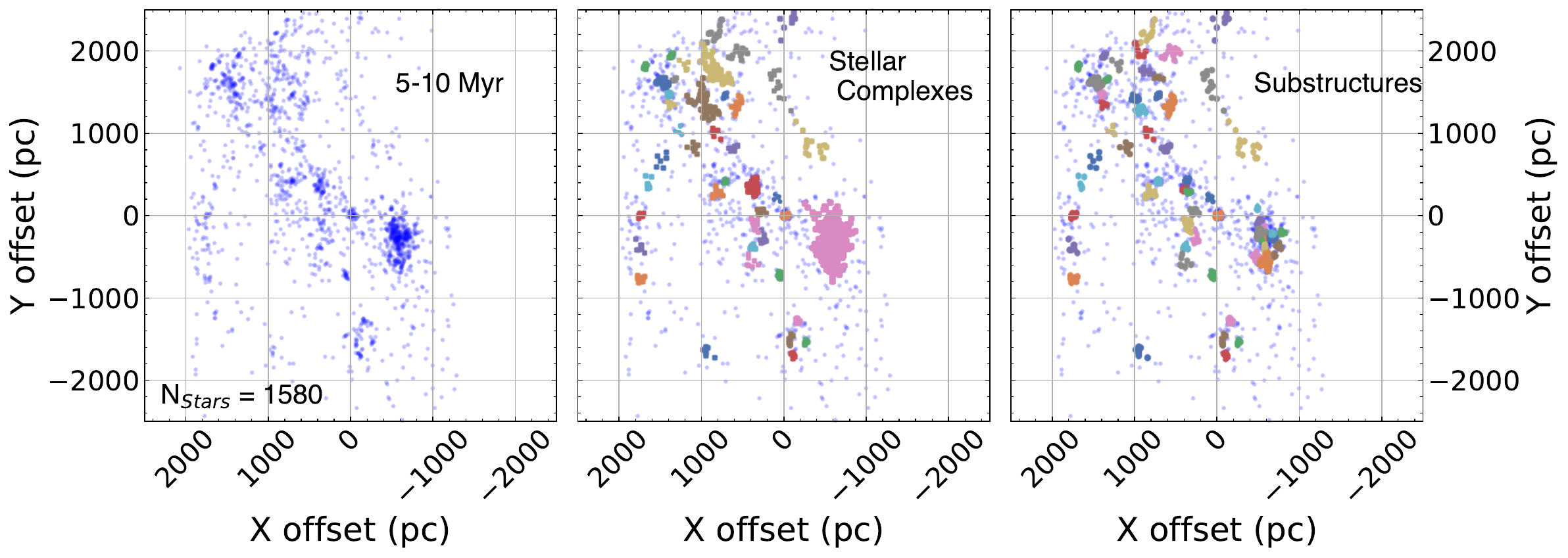}
\includegraphics[width=0.95\textwidth]{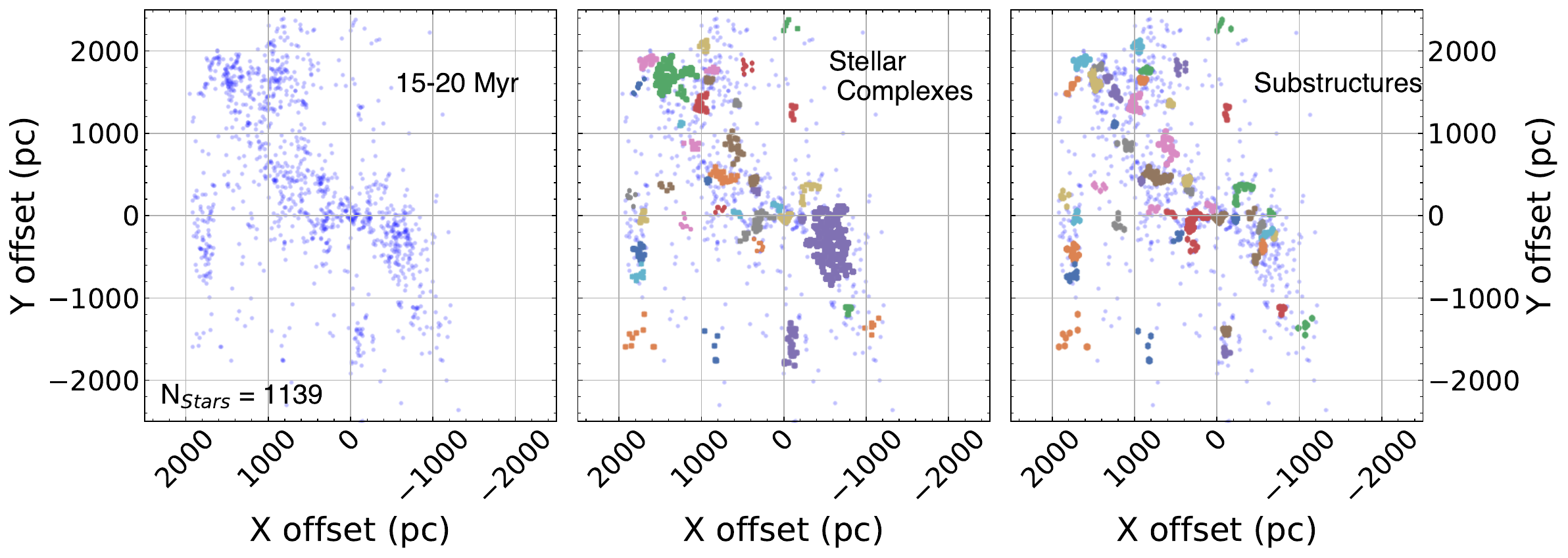}
\caption{The leftmost panel shows the spatial distributions of the field stars in different age groups i.e.  $<$5, 5-10, 10-15, 15-20, 20-30, 30-40 and 40-50 Myr in blue points. The midle and rightmost panels shows the the stellar structures and substructures (respectively) detected in each age group using HDBSCAN. All of the stars in the relevant age group are shown in blue points and the identified structures/substructures  are marked in different colors.}
\label{figapp:age_cluster1}
\end{figure*}

\begin{figure*}[ht!]
\includegraphics[width=0.95\textwidth]{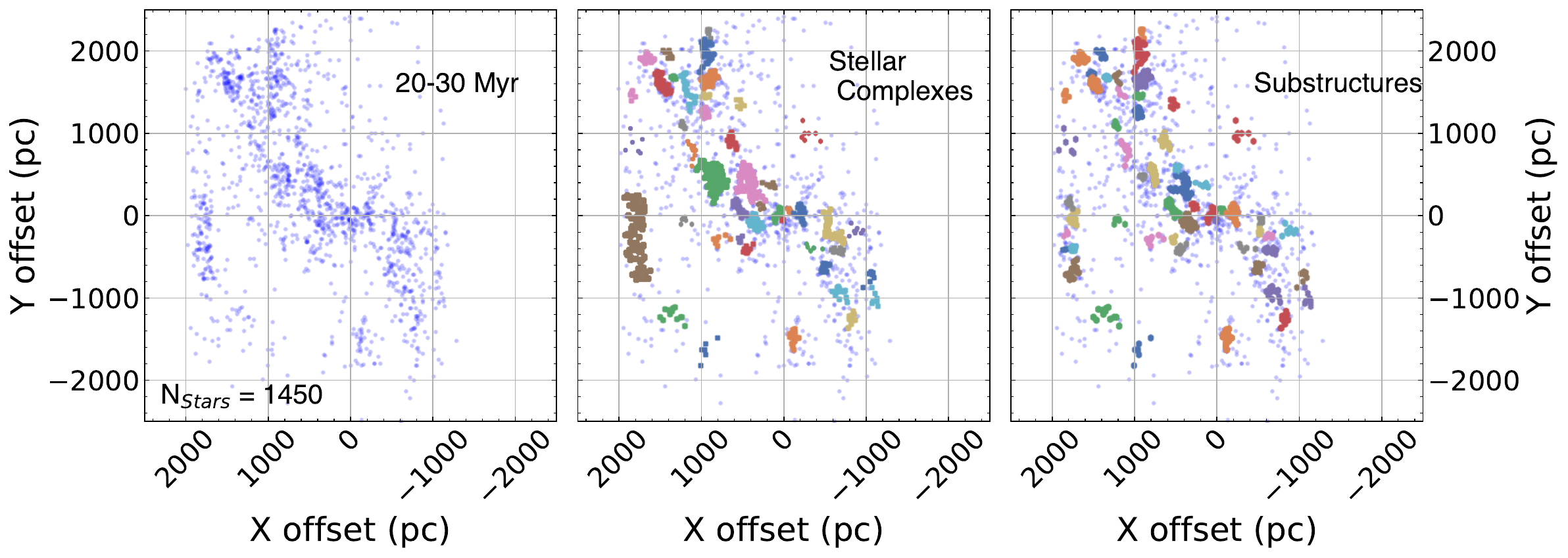}
\includegraphics[width=0.95\textwidth]{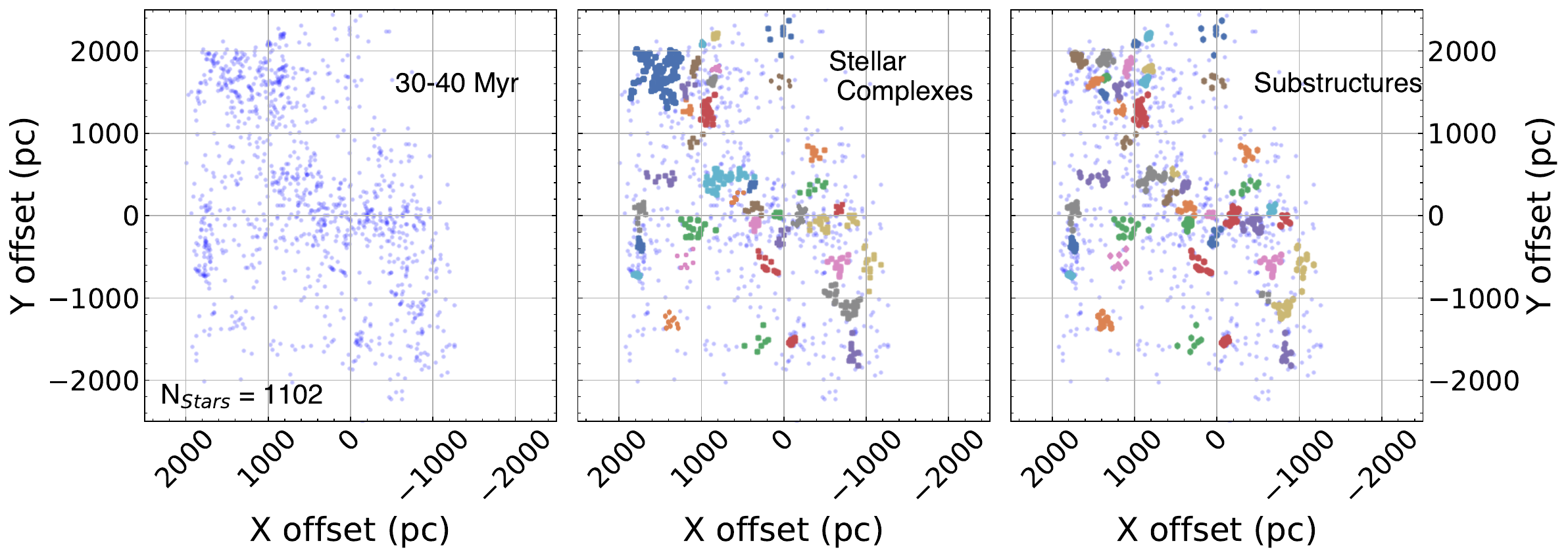}
\includegraphics[width=0.95\textwidth]{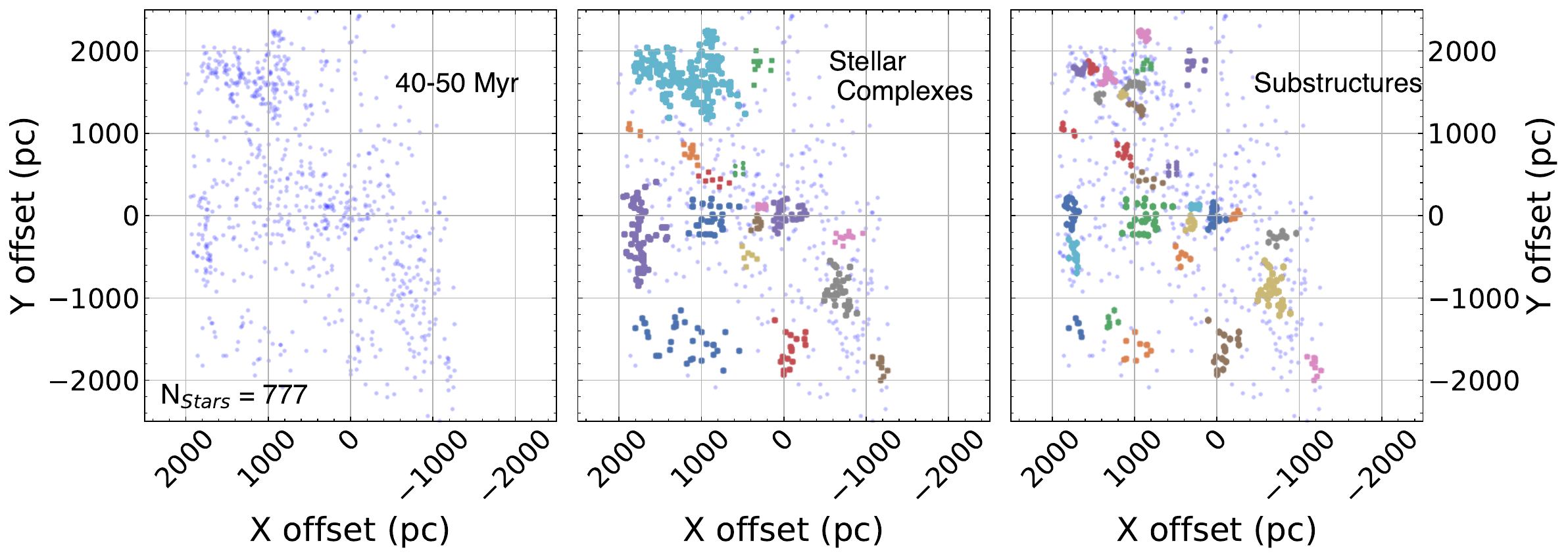}
\caption{Continuation of Figure~\ref{figapp:age_cluster1}}
\label{figapp:age_cluster2}
\end{figure*}

\section{Details of MCMC run}\label{app:mcmc}

We made use of Markov Chain Monte Carlo (MCMC) method within the emcee Python package \citep{Foreman-Mackey2013} to estimate the best-fit parameters for each of the above models (defined by Equations~\ref{eq:pl_exp} and~\ref{eq:pl_exp_const}). Specifically, we estimated the parameters: A, $\theta{_{N}}$ (for model 1) or $\theta{_{N1}}$ (model 2), $\alpha$, $\theta{_{N2}}$ and $\beta$. 
We defined prior distributions for each of these parameter to incorporate prior knowledge and constraints. For instance, we used uniform priors within specified bounds for 0 $<$ A $<$ 50 and 0 $<$ $\alpha$ $<$ 1 and 1 $<$ $\theta_N$ $<$ $\theta_{max}$ and 0 $<$ $\beta$ $<$ 1 for model~1 and same for model~2 with addition of 1 $<$ $\theta{_{N1}}$  $<$ 2 and $\theta{_{N1}}$  $<$ $\theta{_{N2}}$  $<$ $\theta_{max}$.   
These priors ensured that the parameters remained within physically plausible ranges.

The MCMC sampling was conducted using 50 walkers, with each walker initialized at positions sampled from the prior distributions. We performed 5,000 iterations to ensure sufficient exploration of the parameter space. To account for potential autocorrelation and to allow the chains to converge to the posterior distribution, we used a 100-iteration burn-in period, during which 100 samples were discarded. The remaining samples were used to estimate the posterior distribution of the parameters. The best-fit values were computed as the median of the posterior samples, along with their associated uncertainties as median absolute deviation.

To compare the different models, we used statistical metrics to assess their relative performance. Specifically, we employed the Bayesian Information Criterion (BIC), which balances the model's goodness of fit against its complexity. 
We computed the BIC for each model based on the likelihood of the data given the model and the number of parameters. The model with the lowest BIC value was considered the best fit to the data, as it achieved a good fit while avoiding unnecessary complexity.

\end{document}